\documentclass{iopjournal}

\usepackage{amsmath}
\usepackage[T1]{fontenc}
\usepackage{lmodern}
\usepackage[numbers,sort&compress]{natbib}
\usepackage{tikz}
\usepackage{booktabs}
\usepackage{multirow}
\usetikzlibrary{decorations.pathreplacing}
\usepackage{hyperref}
\usepackage[capitalize]{cleveref}

\newcommand{\OJA}{{\mbox{\normalfont\scshape OmniJet}-$\alpha$}}
\newcommand{\OJAC}{\mbox{{\normalfont\scshape OmniJet}-$\alpha_C$}}
\newcommand{\GH}{{\normalfont\scshape GettingHigh}}
\newcommand{\GS}{{\normalfont\scshape GettingSquare}}
\newcommand{\AS}{{\normalfont\scshape AllShowers}}

\begin{document}

\title{\boldmath SPADE: Split-and-Delay Embeddings for Autoregressive High-Granularity Calorimeter Simulation}

\author{Joschka Birk$^1$\orcid{0000-0002-1931-0127}, 
    Frank Gaede$^2$\orcid{0000-0002-7055-9200},
    Anna Hallin$^1$\orcid{0000-0002-1551-814X}, 
    Gregor Kasieczka$^1$\orcid{0000-0003-3457-2755},
    Martina Mozzanica$^1$\orcid{0009-0002-1111-6247},
    Henning Rose$^{1,*}$\orcid{0009-0002-2062-5246}
    }

\affil{$^1$Institute for Experimental Physics, Universität Hamburg,
	Luruper Chaussee 149, 22761 Hamburg, Germany}

\affil{$^2$Deutsches Elektronen-Synchrotron DESY,
    Notkestr. 85, 22607 Hamburg, Germany}

\affil{$^*$Author to whom any correspondence should be addressed.}

\email{henning.rose@uni-hamburg.de}


\begin{abstract}

We introduce SPADE (SPlit And Delay Embeddings), an autoregressive transformer for sequences whose tokens carry multiple features. Rather than embedding these features jointly, SPADE embeds them independently. Delaying each feature stream relative to the previous one allows intra-token correlations to be learned by the standard self-attention mechanism. Applied to point-cloud calorimeter shower generation in the highly granular ILD detector, SPADE is competitive with the state of the art \AS{} model on photon showers, and substantially outperforms its VQ-VAE-based predecessor \OJAC{}. The mechanism is applicable to any generative task with multi-feature tokens, enabling LLM-style pretraining workflows for higher-dimensional data.

\end{abstract}

\section{Introduction}
Current and future collider experiments in high-energy physics (HEP) require an increasing amount of Monte Carlo (MC) samples. Tools like GEANT4~\cite{Agostinelli:2002hh} produce high-fidelity detector simulations, but require large amounts of compute resources, particularly for calorimeters. It is expected that the available resources will not be able to keep up with the demand~\cite{2019,Boehnlein:2803119}. To avoid this becoming a bottleneck, there is a high interest in the development of generative machine learning (ML) methods that can supplement the MC samples. Existing methods include generative adversarial networks (GANs)~\cite{Paganini:2017hrr, Paganini:2017dwg, deOliveira:2017rwa, Erdmann:2018kuh, Erdmann:2018jxd, Musella:2018rdi, Belayneh:2019vyx, Butter:2020qhk, ATLAS:2020, Ghosh:2020kkt, ATLAS:2021pzo, ATLAS:2022jhk, FaucciGiannelli:2023fow, Dogru:2024gpk}, variational autoencoders (VAEs)~\cite{ATLAS:2022jhk, Cresswell:2022tof, Hoque:2023zjt, Liu:2024kvv} and diffusion and flow models~\cite{Krause:2021ilc, Krause:2021wez, Schnake:2022, Krause:2022jna, Xu:2023xdc, Buckley:2023daw, Pang:2023wfx, Ernst:2023qvn, Schnake:2024mip, Du:2024gbp, Majerz:2025ykn, Mikuni:2022xry, Acosta:2023zik, Mikuni:2023tqg, Amram:2023onf, Jiang:2024ohg, Kobylianskii:2024ijw, Jiang:2024bwr, Favaro:2024rle, hildebrandt2026brickscompositionalneuralmarkov, li2026nestedgptvariablemultiplicitypartonshowers}.

While most research into generative ML has focused on dedicated models, foundation models (FM)~\cite{bommasani2022opportunitiesrisksfoundationmodels} with generative capabilities could provide another opportunity. %
Most foundation models built for HEP so far focus on jets and use some type of fill-in-the-blank pretraining target~\cite{Hallin:2025ywf}. The model \OJA~\cite{Birk:2024knn}, originally developed for jets, uses a next token prediction target similar to GPT~\cite{Radford2018ImprovingLU}. This means that the pretraining itself results in a generative model. Since the input features are continuous, a VQ-VAE~\cite{oord2018neural,bao2022beit,huh2023straightening} 
converts them into discrete tokens to serve as targets for next token prediction.
We have previously successfully adapted \OJA{} to generate calorimeter showers, resulting in the model \OJAC~\cite{Birk:2025wai}. Instead of the jet constituent features of \OJA, the input to this model consists of the spatial coordinates $(x,y,z)$ of the shower deposits, as well as the deposited energy $E$. Just as \OJA, \OJAC{} uses a VQ-VAE to tokenize the input features, replacing each $(x, y, z, E)$ input vector with a single integer (token). While this is a straightforward approach following the initial implementation of \OJA, it is not the most effective. In contrast to the fully continuous features of the jet constituents of the original model, which can't immediately be used as targets for next token prediction, three of the four calorimeter features are discrete. It should therefore in principle be possible to use these as-is. 
Furthermore, it is difficult to scale \OJAC{} to higher granularities: the higher the resolution needs to be, the more tokens the VQ-VAE needs. This affects not only the VQ-VAE itself, but also the main model, as the tokens need to be mapped to the embedding space and then back. Increasing the number of possible tokens in this mapping leads to a steep rise in the number of model parameters, resulting in increased training and inference times.

This work introduces SPADE: Split And Delay Embeddings for arbitrary detector granularities. SPADE \emph{splits} the four hit features (three spatial coordinates and the deposited energy) into independent prediction streams, replacing the single VQ-VAE token of \OJAC{}. This has two benefits: first, eliminating the VQ-VAE also eliminates the information loss inherent to VQ-VAE;
second, using four separate prediction streams reduces the spatial vocabulary from cubic to linear in the per-axis grid resolution. Predicting the four features independently would, however, lead to the correlations between them being lost. SPADE preserves these by staggering the streams along the sequence with progressively increasing \emph{delays}. Instead of generating a hit all at once, the model builds each hit sequentially, allowing the autoregressive conditioning to maintain the crucial correlations between a hit's features.

Several recent approaches address related tokenization strategies. Ref.~\cite{cardonagiraldo2026generalizablefoundationmodelscalorimetry} also uses autoregressive token generation for calorimeter showers but keeps a 3D voxel index, scaling as $\mathcal{O}(N^3)$. Nested-GPT~\cite{li2026nestedgptvariablemultiplicitypartonshowers} factorizes per-token features but recovers their correlations through an inner recurrent decoder, whereas SPADE uses a delay along the main sequence. A similar delay mechanism has been applied earlier in audio generation to parallel residual-VQ codebooks~\cite{copet2024simplecontrollablemusicgeneration}, meaning that different representations of the same waveform functioned as different feature streams. However, our delayed features are not different representations of the same object, but distinct physical coordinates of a single hit.

The model is evaluated using two photon shower datasets: \GH{} (irregular grid) and \GS{} (regular grid, multiple granularities). In addition to a comparison with \OJAC, we compare the performance of SPADE to a baseline model we refer to as \textit{Combined}. This model removes the need for a VQ-VAE, but uses a single combined spatial vocabulary. We show that SPADE achieves superior performance to both \OJAC{} and the Combined baseline across photon showers. Furthermore, while both SPADE and the Combined baseline eliminate the lossy VQ-VAE of \OJAC{}, SPADE uniquely avoids the Combined model's poor parameter scaling by reducing the parameter count by orders of magnitude. On the irregular \GH{} dataset, SPADE is competitive with the state of the art \AS{}~\cite{Buss:2026yrf} reference model on most observables. On the regular \GS{} dataset, the factorized SPADE embedding shows particularly clear advantages over Combined on observables that probe the spatial structure of the shower. Crucially, SPADE scales linearly rather than cubically with detector granularity, making it a practical architecture for the highly granular calorimeters of future collider experiments.

This paper is organized as follows. \cref{sec:dataset} describes the dataset used for training and evaluation. \cref{sec:methods} introduces the three autoregressive architectures compared in this work: \OJAC, Combined, and SPADE, as well as the flow matching \AS{} reference model. \cref{sec:results} presents the results: a four-way model comparison on the \GH{} dataset with a realistic, irregular cell geometry (\cref{sec:results_gettinghigh}); and an evaluation of physics performance and parameter scaling on the regular \GS{} dataset (\cref{sec:results_gettingsquare}). Additionally we show the superior training and representation efficiency in \cref{sec:training_efficiency}, and evaluate the inference time in \cref{sec:inference_time}. \cref{sec:conclusions} summarizes the findings and discusses the broader applicability of the split-and-delay mechanism.

\section{Dataset}
\label{sec:dataset}

All photon datasets used in this work derive from the same Geant4~\cite{Agostinelli:2002hh} simulation campaign, originally produced for Ref.~\cite{Buhmann:2020pmy} using a detailed DD4hep~\cite{Frank_2014} model of the Si-W electromagnetic calorimeter of the International Large Detector (ILD)~\cite{ILDConceptGroup:2020sfq, Suehara:2018mqk}. Single photons with energies uniformly distributed between 10 and 100~GeV enter the calorimeter perpendicularly at $z=0$ and travel along the $z$-axis. 

Geant4 records each individual energy deposition as a \emph{step}; a \emph{hit} is the sum of all steps that fall into a single voxel of a chosen detector grid. We use two datasets, \GH{} and \GS{}, that share the same underlying Geant4 steps but bin them onto different voxel grids.

\begin{figure}[t]
\centering
\definecolor{seabornblue}{HTML}{4C72B0}

\newcommand{\edot}[4]{%
  \pgfmathsetmacro{\dsize}{3.2*sqrt(#2)}%
  \node[circle, fill=#1, draw=none, inner sep=0pt,
        minimum size=\dsize pt] at ({#3},{#4}) {};}

\begin{tikzpicture}[
    panel/.style      = {draw=black, line width=0.8pt},
    gridline/.style   = {draw=black!22, line width=0.3pt},
    badge/.style      = {font=\scriptsize, text=black!60},
    paneltitle/.style = {font=\bfseries\small, align=center},
    panelsub/.style   = {font=\scriptsize, text=black!55, align=center},
    flow/.style       = {->, >=stealth, thick, draw=black!70},
    flowlab/.style    = {font=\scriptsize, text=black!70, align=center}
]
\pgfmathsetmacro{\psize}{2.4}
\pgfmathsetmacro{\sC}{4.2}

\begin{scope}
    \draw[panel] (0,0) rectangle (\psize,\psize);
    \edot{seabornblue!80}{1.0}{0.83*\psize}{0.95*\psize}
    \edot{seabornblue!80}{0.9}{0.55*\psize}{0.82*\psize}
    \edot{seabornblue!80}{0.6}{0.70*\psize}{0.79*\psize}
    \edot{seabornblue!80}{1.2}{0.55*\psize}{0.55*\psize}
    \edot{seabornblue!80}{0.8}{0.71*\psize}{0.57*\psize}
    \edot{seabornblue!80}{1.3}{0.56*\psize}{0.30*\psize}
    \edot{seabornblue!80}{0.7}{0.31*\psize}{0.33*\psize}
    \node[badge, anchor=north west] at ({0.04*\psize},{0.965*\psize}) {7 steps};
    \node[panelsub,   anchor=south] at ({0.5*\psize},{\psize+0.05}) {continuous deposits};
    \node[paneltitle, anchor=south] at ({0.5*\psize},{\psize+0.35}) {Geant4 steps};
\end{scope}

\draw[flow] ({\psize+0.1},{0.5*\psize}) -- ({\sC-0.1},{0.5*\psize});
\node[flowlab] at ({(\psize+\sC)/2},{0.5*\psize+0.32}) {map to\\staggered grid};

\begin{scope}[shift={(\sC,0)}]
    \draw[gridline] (0,{0.5*\psize})--(\psize,{0.5*\psize});
    \draw[gridline] ({0.5*\psize},0)--({0.5*\psize},{0.5*\psize});
    \draw[gridline] ({0.25*\psize},{0.5*\psize})--({0.25*\psize},\psize);
    \draw[gridline] ({0.75*\psize},{0.5*\psize})--({0.75*\psize},\psize);
    \draw[panel] (0,0) rectangle (\psize,\psize);
    \edot{seabornblue}{0.7}{0.25*\psize}{0.25*\psize}   %
    \edot{seabornblue}{1.3}{0.75*\psize}{0.25*\psize}   %
    \edot{seabornblue}{1.0}{0.875*\psize}{0.75*\psize}  %
    \edot{seabornblue}{3.5}{0.5*\psize}{0.75*\psize}
    \node[badge, anchor=north west] at ({0.04*\psize},{0.965*\psize}) {4 hits};
    \node[panelsub,   anchor=south] at ({0.5*\psize},{\psize+0.05}) {30$\times$30$\times$30, irregular};
    \node[paneltitle, anchor=south] at ({0.5*\psize},{\psize+0.35}) {GettingHigh};
\end{scope}
\end{tikzpicture}
\caption{Mapping of continuous Geant4 shower deposits (left) to the \GH{} representation (right). While the original Geant4 steps record exact spatial coordinates, \GH{} bins these continuous deposits into a nominal $30\times30\times30$ geometry by mapping them onto physical, layer-staggered ILD sensors (visible as the offset between the upper and lower row dividers). Marker size in the \GH{} panel scales with the per-sensor summed energy.}
\label{fig:gh_vs_x1}
\end{figure}
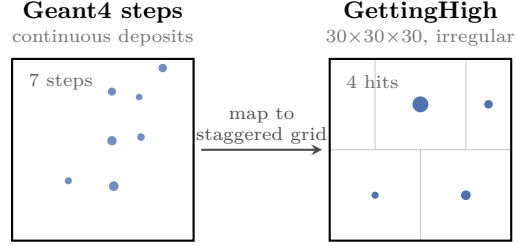

\paragraph{\GH\ (irregular baseline).}
The \GH{} dataset maps each Geant4 step onto its physical sensor in a $30 \times 30 \times 30$ array; each non-empty sensor becomes one hit. Because the realistic ECal barrel prototype has slightly staggered layers, this one-to-one mapping inherits the irregular geometry of the real detector rather than imposing a regular grid, as illustrated in \cref{fig:gh_vs_x1}. The dataset contains 950\,000 showers, split into 760\,000 for training, 95\,000 for validation and 95\,000 for testing.

\begin{figure}[t]
\centering
\resizebox{\linewidth}{!}{\definecolor{seabornblue}{HTML}{4C72B0}

\providecommand{\edot}[4]{%
  \pgfmathsetmacro{\dsize}{3.2*sqrt(#2)}%
  \node[circle, fill=#1, draw=none, inner sep=0pt,
        minimum size=\dsize pt] at ({#3},{#4}) {};}

\begin{tikzpicture}[
    panel/.style      = {draw=black, line width=0.8pt},
    gridline/.style   = {draw=black!22, line width=0.3pt},
    count/.style      = {font=\tiny\bfseries, text=black!75, inner sep=1pt},
    paneltitle/.style = {font=\bfseries\small, align=center},
    panelsub/.style   = {font=\scriptsize, text=black!55, align=center},
    flow/.style       = {->, line width=0.8pt, draw=black!55},
    flowlab/.style    = {font=\scriptsize, text=black!60, align=center}
]
\pgfmathsetmacro{\psize}{2.4}
\pgfmathsetmacro{\pitch}{3.8}
\pgfmathsetmacro{\sB}{\pitch}
\pgfmathsetmacro{\sC}{2*\pitch}
\pgfmathsetmacro{\sD}{3*\pitch}
\pgfmathsetmacro{\sE}{4*\pitch}

\begin{scope}
    \draw[panel] (0,0) rectangle (\psize,\psize);
    \edot{seabornblue!80}{1.0}{0.83*\psize}{0.95*\psize}
    \edot{seabornblue!80}{0.9}{0.55*\psize}{0.82*\psize}
    \edot{seabornblue!80}{0.6}{0.70*\psize}{0.79*\psize}
    \edot{seabornblue!80}{1.2}{0.55*\psize}{0.55*\psize}
    \edot{seabornblue!80}{0.8}{0.71*\psize}{0.57*\psize}
    \edot{seabornblue!80}{1.3}{0.56*\psize}{0.30*\psize}
    \edot{seabornblue!80}{0.7}{0.31*\psize}{0.33*\psize}
    \node[panelsub,   anchor=south] at ({0.5*\psize},{\psize+0.05}) {continuous deposits};
    \node[paneltitle, anchor=south] at ({0.5*\psize},{\psize+0.33}) {Geant4 steps};
\end{scope}

\begin{scope}[shift={(\sB,0)}]
    \foreach \i in {1,...,7} {%
        \pgfmathsetmacro{\xc}{\i*\psize/8}
        \draw[gridline] (\xc,0)--(\xc,\psize); \draw[gridline] (0,\xc)--(\psize,\xc);}
    \draw[panel] (0,0) rectangle (\psize,\psize);
    \edot{seabornblue}{1.0}{0.8125*\psize}{0.9375*\psize}
    \edot{seabornblue}{0.9}{0.5625*\psize}{0.8125*\psize}
    \edot{seabornblue}{0.6}{0.6875*\psize}{0.8125*\psize}
    \edot{seabornblue}{1.2}{0.5625*\psize}{0.5625*\psize}
    \edot{seabornblue}{0.8}{0.6875*\psize}{0.5625*\psize}
    \edot{seabornblue}{1.3}{0.5625*\psize}{0.3125*\psize}
    \edot{seabornblue}{0.7}{0.3125*\psize}{0.3125*\psize}
    \node[panelsub,   anchor=south] at ({0.5*\psize},{\psize+0.05}) {$120\times120\times30$};
    \node[paneltitle, anchor=south] at ({0.5*\psize},{\psize+0.36}) {x16};
\end{scope}

\begin{scope}[shift={(\sC,0)}]
    \foreach \i in {1,...,3} {%
        \pgfmathsetmacro{\xc}{\i*\psize/4}
        \draw[gridline] (\xc,0)--(\xc,\psize); \draw[gridline] (0,\xc)--(\psize,\xc);}
    \draw[panel] (0,0) rectangle (\psize,\psize);
    \edot{seabornblue}{1.0}{0.875*\psize}{0.875*\psize}
    \edot{seabornblue}{0.9}{0.580*\psize}{0.875*\psize}
    \edot{seabornblue}{0.6}{0.670*\psize}{0.875*\psize}
    \node[count] at ({0.625*\psize},{0.955*\psize}) {2};
    \edot{seabornblue}{1.2}{0.580*\psize}{0.625*\psize}
    \edot{seabornblue}{0.8}{0.670*\psize}{0.625*\psize}
    \node[count] at ({0.625*\psize},{0.705*\psize}) {2};
    \edot{seabornblue}{1.3}{0.625*\psize}{0.375*\psize}
    \edot{seabornblue}{0.7}{0.375*\psize}{0.375*\psize}
    \node[panelsub,   anchor=south] at ({0.5*\psize},{\psize+0.05}) {$60\times60\times30$};
    \node[paneltitle, anchor=south] at ({0.5*\psize},{\psize+0.36}) {x4};
\end{scope}

\begin{scope}[shift={(\sD,0)}]
    \draw[gridline] ({0.5*\psize},0)--({0.5*\psize},\psize);
    \draw[gridline] (0,{0.5*\psize})--(\psize,{0.5*\psize});
    \draw[panel] (0,0) rectangle (\psize,\psize);
    \edot{seabornblue}{1.2}{0.75*\psize}{0.75*\psize}
    \edot{seabornblue}{0.9}{0.66*\psize}{0.75*\psize}
    \edot{seabornblue}{1.0}{0.84*\psize}{0.75*\psize}
    \edot{seabornblue}{0.8}{0.75*\psize}{0.66*\psize}
    \edot{seabornblue}{0.6}{0.75*\psize}{0.84*\psize}
    \node[count] at ({0.75*\psize},{0.95*\psize}) {5};
    \edot{seabornblue}{1.3}{0.75*\psize}{0.25*\psize}
    \edot{seabornblue}{0.7}{0.25*\psize}{0.25*\psize}
    \node[panelsub,   anchor=south] at ({0.5*\psize},{\psize+0.05}) {$30\times30\times30$};
    \node[paneltitle, anchor=south] at ({0.5*\psize},{\psize+0.36 }) {x1};
\end{scope}

\begin{scope}[shift={(\sE,0)}]
    \draw[gridline] ({0.5*\psize},0)--({0.5*\psize},\psize);
    \draw[gridline] (0,{0.5*\psize})--(\psize,{0.5*\psize});
    \draw[panel] (0,0) rectangle (\psize,\psize);
    \edot{seabornblue}{4.5}{0.75*\psize}{0.75*\psize}
    \edot{seabornblue}{1.3}{0.75*\psize}{0.25*\psize}
    \edot{seabornblue}{0.7}{0.25*\psize}{0.25*\psize}
    \node[panelsub,   anchor=south] at ({0.5*\psize},{\psize+0.05}) {$30\times30\times30$};
    \node[paneltitle, anchor=south] at ({0.5*\psize},{\psize+0.30}) {x1 merged};
\end{scope}

\draw[flow] ({\psize},{0.5*\psize}) -- ({\sB},{0.5*\psize});
\node[flowlab] at ({(\psize+\sB)/2},{0.5*\psize+0.32}) {bin to\\fine grid};
\draw[flow] ({\sB+\psize},{0.5*\psize}) -- ({\sC},{0.5*\psize});
\node[flowlab] at ({(\sB+\psize+\sC)/2},{0.5*\psize+0.32}) {re-index\\no merge};
\draw[flow] ({\sC+\psize},{0.5*\psize}) -- ({\sD},{0.5*\psize});
\node[flowlab] at ({(\sC+\psize+\sD)/2},{0.5*\psize+0.32}) {re-index\\no merge};
\draw[flow] ({\sD+\psize},{0.5*\psize}) -- ({\sE},{0.5*\psize});
\node[flowlab] at ({(\sD+\psize+\sE)/2},{0.5*\psize+0.32}) {sum\\duplicates};
\draw[decorate, decoration={brace, amplitude=12pt}, line width=0.8pt] 
    ({\sB}, {\psize+0.65}) -- ({\sE+\psize}, {\psize+0.65}) 
    node[midway, above=12pt, font=\bfseries\normalsize] {GettingSquare};
\end{tikzpicture}}
\caption{Construction of the \GS{} datasets. The same Geant4 energy deposits (steps; leftmost panel) are binned onto a fine regular $120\times120\times30$ grid (\GS-x16). The coarser \GS-x4 and \GS-x1 representations are obtained by reassigning each hit to its enclosing voxel at lower transverse resolution, without merging, so all three share an identical hit list, but several hits may occupy the same coarse voxel (numerical labels give the number of duplicate hits per shared cell). The \GS-x1 merged-hit variant additionally sums the energies of duplicate hits into a single hit per cell (enlarged marker).}
\label{fig:gs_pipeline}
\end{figure}
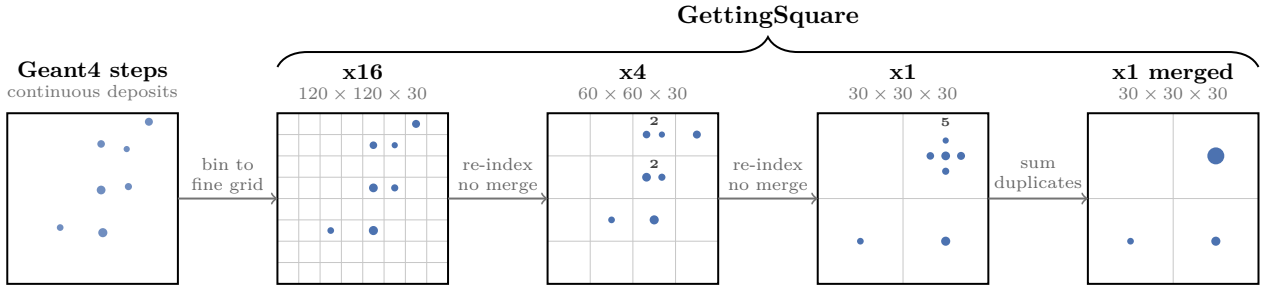

\paragraph{\GS\ (regular grid variants).}
To study the impact of transverse granularity in a controlled setting, we first bin the Geant4 steps onto the finest grid, \textbf{x16} ($120\times120\times30$). The coarser variants are then obtained by reassigning each x16 hit to its enclosing voxel on a $60\times60\times30$ (\textbf{x4}) or $30\times30\times30$ (\textbf{x1}) grid, without merging, so a single coarse voxel can hold multiple hits inherited from x16. The longitudinal segmentation is fixed at 30 layers throughout, isolating transverse granularity as the single experimental variable. This construction is summarized in \cref{fig:gs_pipeline}. Note that \GS-x1 and \GH{} share the same nominal voxel count but differ in geometry: \GS-x1 imposes a regular grid, while \GH{} preserves the irregular detector layout. We additionally use an \textbf{x1-merged} variant in which hits sharing a voxel are summed, mimicking a detector with coarser readout cells. Each \GS{} dataset contains 750\,000 showers, split 495\,000 / 130\,000 / 125\,000 for training, validation and testing. 

\section{Methods}
\label{sec:methods}

The autoregressive models considered in this work, \OJAC, Combined and SPADE, are all decoder-only transformers that act on showers represented as sequences of hits. We define a shower as a collection of $n$ hits, $H = \{h_1, \dots, h_n\}$, where each hit $h_i$ is characterized by its discrete spatial detector coordinates $(x_i, y_i, z_i)$ and its continuous deposited energy $E_i$. The sequence is zero-padded to a maximum length. After mapping the hits to a sequence of tokens, a causal mask is applied so that a token at position $i$ can only attend to tokens at positions $j \le i$, allowing parallel training on next token prediction. At inference, showers are generated autoregressively, one hit at a time. The three models differ in (i) how the hits are ordered, (ii) how each hit is mapped to one or more tokens, and (iii) how the end of the sequence is determined, as detailed below. A schematic comparison of the three architectures is shown in \cref{fig:Model_comparison}.

\subsection{\texorpdfstring{\OJAC}{OJAC}}
\label{sec:OJAC1}
\OJAC{} uses a frozen VQ-VAE to map each four-dimensional hit $(x_i, y_i, z_i, E_i)$ to a single integer token drawn from a learned codebook of size $V = 65\,536$. The transformer is trained as a $V$-way classifier on next token prediction.
After generation, the VQ-VAE decodes the tokens back to physical hit features.

The hits are ordered by deposited energy in descending order. Once tokenized, a special \textit{start token} is added to the start of the sequence, and a \textit{stop token} to the end. During generation, the model will sample autoregressively until it either generates the stop token or reaches the maximum sequence length. 

Two limitations of this design motivate the present work. First, the embedding and unembedding layers contribute $2 \cdot V \cdot d$ parameters, where $d$ is the embedding dimension. With $V$ = 65,536 and $d$ = 256, these two layers alone account for 33.5M of the model's 36M parameters. This increase of the model size leads to slow training times, and scaling $V$ up further to accommodate finer detector granularity quickly becomes prohibitive. Second, the lossy compression of the VQ-VAE leads to spatial and energetic artifacts. Both Combined and SPADE address these issues by replacing the VQ-VAE with a direct embedding scheme, as outlined below.

\subsection{Combined}
\label{sec:combined}
Since the spatial coordinates are already on a discrete grid, it should in principle be possible to use these coordinates directly without tokenization. The Combined model thus removes the VQ-VAE while still using a single \textit{combined} spatial vocabulary. The detector volume is treated as a regular grid of size $N_x \times N_y \times N_z$, and each hit position is mapped to a single scalar ID,
\begin{equation}
\mathrm{ID} = N_y N_z \, x + N_z \, y + z.
\end{equation}
The vocabulary therefore scales multiplicatively with detector granularity, $V = N_x \cdot N_y \cdot N_z$. Unlike \OJAC, energy is not encoded in the spatial token but treated as a separate continuous input. Instead of having a single token prediction head, Combined has one head for predicting the spatial ID, and a mixture-of-Gaussians head (\cref{subsec:energyhead}) for predicting the continuous hit energy (see also \cref{fig:Model_comparison}). The continuous hit energy $E_i$ and the global incident energy $E_{\mathrm{inc}}$ are each passed through a linear projection to 16-dimensional vectors, whereas the dimensionality of the spatial ID embedding is 256. The three components are concatenated and projected to the 256-dimensional input expected by the transformer.  

The spatial ID and the hit energy of the same hit are correlated.
To preserve this correlation while still allowing the spatial and energy heads to predict their respective quantities independently, we delay the energy projection by one step relative to the spatial ID. The input at sequence position $i$ is therefore
\begin{equation}
T_i = \mathrm{Concat}\!\left(\mathrm{Emb}(\mathrm{ID}_i),\, \mathrm{Proj_1}(E_{i-1}),\, \mathrm{Proj_2}(E_{\mathrm{inc}})\right),
\label{eq:combined_token}
\end{equation}
so that by the time the model is about to predict $E_i$, it has already seen (and therefore becomes conditioned on) $\mathrm{ID}_i$.  
The empty slots created by the delay are padded as follows: missing spatial IDs are filled with a dedicated \textit{missing ID} embedding, while missing energy projections are filled with zeros. These padding positions are masked out of the loss so that they do not contribute a training signal. The incident energy $E_{\mathrm{inc}}$ is concatenated to every token, as we found this to strengthen the conditioning and improve fidelity.

Hits are ordered first by longitudinal layer ($z$ ascending), and within each layer by deposited energy ($E$ descending). This layer-by-layer ordering provides a physically meaningful inductive bias that mirrors how showers actually develop. 
The layer-first ordering is used consistently across all Combined and SPADE runs.

\subsection{SPADE}
\label{sec:spade}
\begin{figure}
    \centering
    \includegraphics[width=1\linewidth]{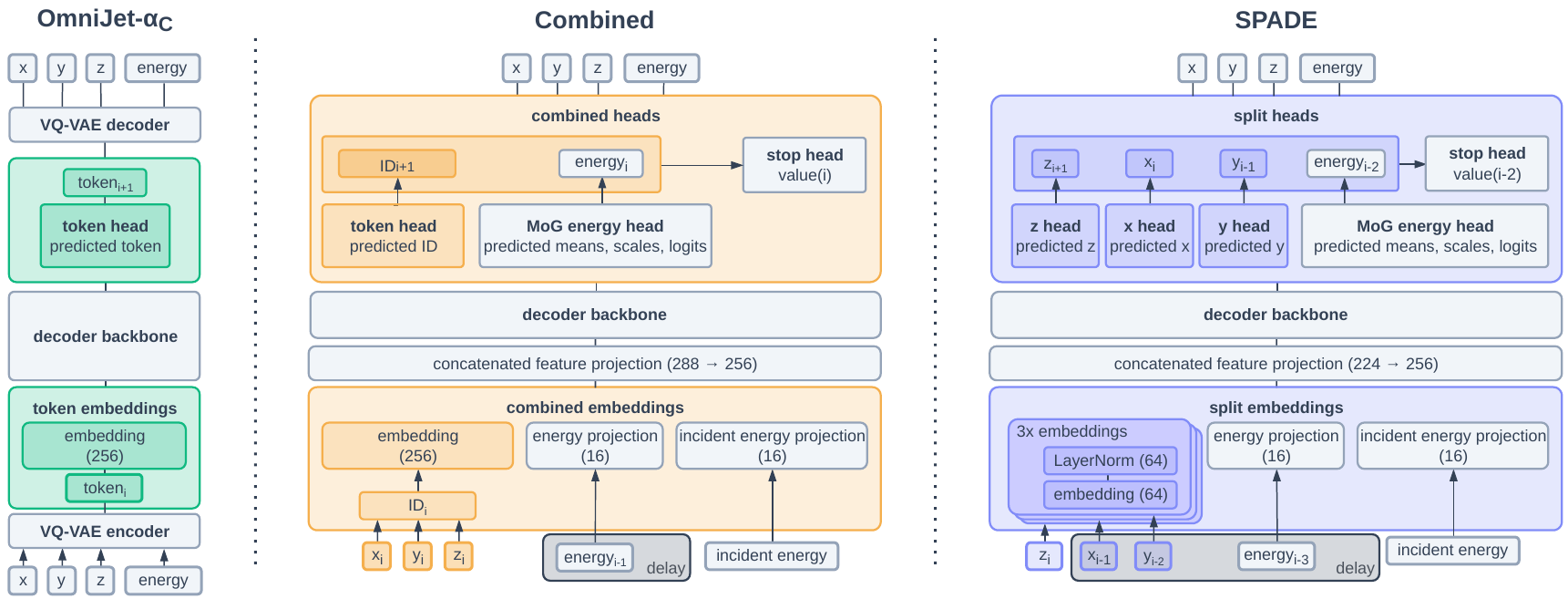}
    \caption{Schematic comparison of the model architectures. \textbf{Left:} \OJAC{} utilizes a VQ-VAE for continuous-to-discrete tokenization of spatial and energy features. \textbf{Middle:} The Combined baseline embeds joint 3D spatial coordinates into a single ID, delaying only the hit energy ($E_{i-1}$) to condition the next prediction. \textbf{Right:} SPADE replaces the joint vocabulary with factorized, independent spatial embeddings stabilized by LayerNorm. SPADE generalizes the delay mechanism by staggering all components ($z_i$, $x_{i-1}$, $y_{i-2}$, $E_{i-3}$) to enforce autoregressive conditioning among intra-hit components. Both Combined and SPADE utilize a Mixture of Gaussians (MoG) head for continuous energy prediction alongside a stop head for sequence termination.}
    \label{fig:Model_comparison}
\end{figure}

SPADE replaces the single combined vocabulary with a \textit{factorized} one. Rather than embedding a 3D voxel index, the three spatial coordinates $x$, $y$, and $z$ are embedded independently into 64-dimensional vectors, reducing the vocabulary scaling from multiplicative to additive: $V = N_x + N_y + N_z$. The single spatial ID prediction head from Combined now becomes three separate heads, one for each spatial coordinate, as illustrated in \cref{fig:Model_comparison}. We found that without normalization, the relative scale of the three spatial embeddings drifted during training and destabilized convergence. We therefore apply a LayerNorm~\cite{ba2016layernormalization} independently to each spatial embedding before they are combined. The continuous hit energy is predicted by the same mixture-of-Gaussians head as in Combined, described in \cref{subsec:energyhead}.

SPADE generalizes the delay mechanism of Combined to the factorized setting. 
We stagger all four components ($x$, $y$, $z$, $E$) along the sequence with progressively increasing delays, so that the input at sequence position $i$ contains components belonging to four different hits:
\begin{equation}
T_i = \mathrm{Concat}\!\left(\mathrm{Emb_z}(z_i),\, \mathrm{Emb_x}(x_{i-1}),\, \mathrm{Emb_y}(y_{i-2}),\, \mathrm{Proj_1}(E_{i-3}),\, \mathrm{Proj_2}(E_{\mathrm{inc}})\right).
\label{eq:spade_token}
\end{equation}
Each coordinate of a hit is now predicted at a different sequence position, so the heads condition on each other implicitly through the autoregressive structure. By the time $x_i$ is to be predicted, it has already seen (and therefore becomes conditioned on) $z_i$. In the same way, $y_i$ is conditioned on $z_i$ and $x_i$, and $E_i$ is conditioned on all spatial coordinates for position $i$.

The empty slots at the start and end of the sequence are padded and masked out of the loss exactly as in Combined. $E_{\mathrm{inc}}$ is appended to every token, and the concatenated input of \cref{eq:spade_token} is linearly projected to the 256-dimensional transformer input.

\subsection{Energy head}
\label{subsec:energyhead}

Both Combined and SPADE predict the continuous hit energy $E_i$ through mixture-of-Gaussians (MoG) head~\cite{Bishop:1994} that parametrizes its conditional density as a mixture of $K$ Gaussians:
\begin{equation}
    p(E_i) = \sum_{k=1}^{K} \pi_k\, \mathcal{N}(E_i \mid \mu_k, \sigma_k^{2}),
\label{eq:mog}
\end{equation}
where the mixture weights $\pi_k$, means $\mu_k$ and variances $\sigma_k^{2}$ are produced at each sequence position by a small MLP sharing its inputs with the spatial heads. The spatial coordinates of the same hit are already encoded in the transformer's hidden state at that position through the staggered input of \cref{eq:combined_token,eq:spade_token} and the causal self-attention, so the energy prediction is implicitly conditioned on $(x_i, y_i, z_i)$. We use $K=4$ throughout. Because hits within a layer are ordered by descending energy, the energy $E_{i-1}$ is often numerically very close to the target $E_i$. To prevent the model from exploiting this ordering by trivially copying the previous input rather than learning the true distribution, we inject relative Gaussian noise into the continuous energy inputs during training. The head is trained by minimizing the negative log-likelihood of the predicted energy distributions, summed over non-padded positions. At inference, energies are sampled from the predicted mixture.

\subsection{Stop head}
\label{subsec:stophead}
Combined and SPADE could in principle use an \OJAC{}-like \emph{stop token}, but the staggered input scheme makes this awkward to implement consistently. We instead introduce a dedicated stop head, which has the additional benefit of giving both models the same termination mechanism and enabling a clean head-to-head comparison. The stop head is a lightweight binary classifier with two output tokens, \textit{continue} and \textit{stop}, where the former allows shower token generation to proceed and the latter terminates it. Crucially, the stop head does not take the output of the backbone as input, but only the current shower state as predicted by the $x$, $y$, $z$, and $E$ heads. Incorporating the backbone output was found to significantly degrade results. The stop head is trained in a second stage with the backbone frozen, after which the final checkpoint is selected from the converged epochs by matching the generated $n_{\text{hits}}$ distribution to Geant4 on the validation set, with priority given to avoiding outlier showers of implausible length. Despite this tuning, we observe a specific failure mode during inference for SPADE where the model occasionally halt generation prematurely, leading to a severe under-prediction of the total deposited energy. This behavior affects around 0.35\% of raw SPADE showers. To mitigate this, sampled showers are filtered through a short postprocessing step to reject these early-stopping failures. This mitigation strategy is described in detail in Appendix~\ref{app:postprocessing}, and all physics results in this paper are reported after this postprocessing is applied.

\subsection{Decoder backbone}
Combined and SPADE utilize the same decoder architecture, featuring several upgrades over the \OJAC{} baseline. Relative sequence positions are encoded using Rotary Position Embedding (RoPE)~\cite{SU2024127063}. To maximize training efficiency and generation throughput, the attention mechanism utilizes Multi-Query Attention~\cite{shazeer2019fast} and hardware-optimized FlashAttention kernels~\cite{dao2022flashattention, dao2023flashattention2}, alongside key-value (KV) caching during autoregressive inference.

\subsection{Training}
Both Combined and SPADE are trained with the Ranger optimizer (Lookahead~\cite{NEURIPS2019_90fd4f88} + RAdam~\cite{Liu2020On}) and constant learning-rate with linear warmup. The total loss combines the cross-entropy or negative log-likelihood loss from each prediction head. We observed that adjacent training checkpoints can differ noticeably in the quality of generated showers even when their validation losses are essentially identical; to suppress this checkpoint-to-checkpoint variability, sampling uses an exponential moving average (EMA)~\cite{tarvainen2018meanteachersbetterrole} of the training weights rather than the weights of any individual checkpoint. The full set of hyperparameters is listed in Appendix~\ref{sec:appendix_hyperparams}.

\subsection{\AS}
\label{subsec:allshowers}
One of the current state of the art approaches for generative modeling of calorimeter showers is the \AS{} framework~\cite{Buss:2026yrf}. This model provides a unified architecture designed to learn shower development across different particle types, incident angles, and incoming energies within a single conditional generative model. \AS{} consists of two models: the PointCountFM, which predicts the number of points per layer and directly determines the $z$ coordinate of each point, and a CNF-transformer, which predicts the remaining features ($x$, $y$ and energy $E$). The $z$ coordinate is deterministically assigned from the layer index, reducing the generative task of the CNF-transformer to three dimensions.

In this work, we train the \AS{} model on the photon datasets described in \cref{sec:dataset} (\GH{} and \GS). 
We remove the conditioning on particle type and incident angle, as both are redundant in this study, leaving the incident energy as the only conditioning feature. 
The input layer is resized accordingly; the rest of the architecture is unchanged.

\section{Results}
\label{sec:results}
In the following we show the results grouped by the datasets the models were trained on. In \cref{sec:results_gettinghigh} we compare \OJAC{}, Combined, SPADE and \AS{} on the irregular \GH{} dataset, whereas \cref{sec:results_gettingsquare} 
shows the performance of Combined, SPADE and \AS{} on the \GS{} dataset.

\subsection{\GH}
\label{sec:results_gettinghigh}
\begin{figure}[t]
    \centering
    \includegraphics[width=1\linewidth]{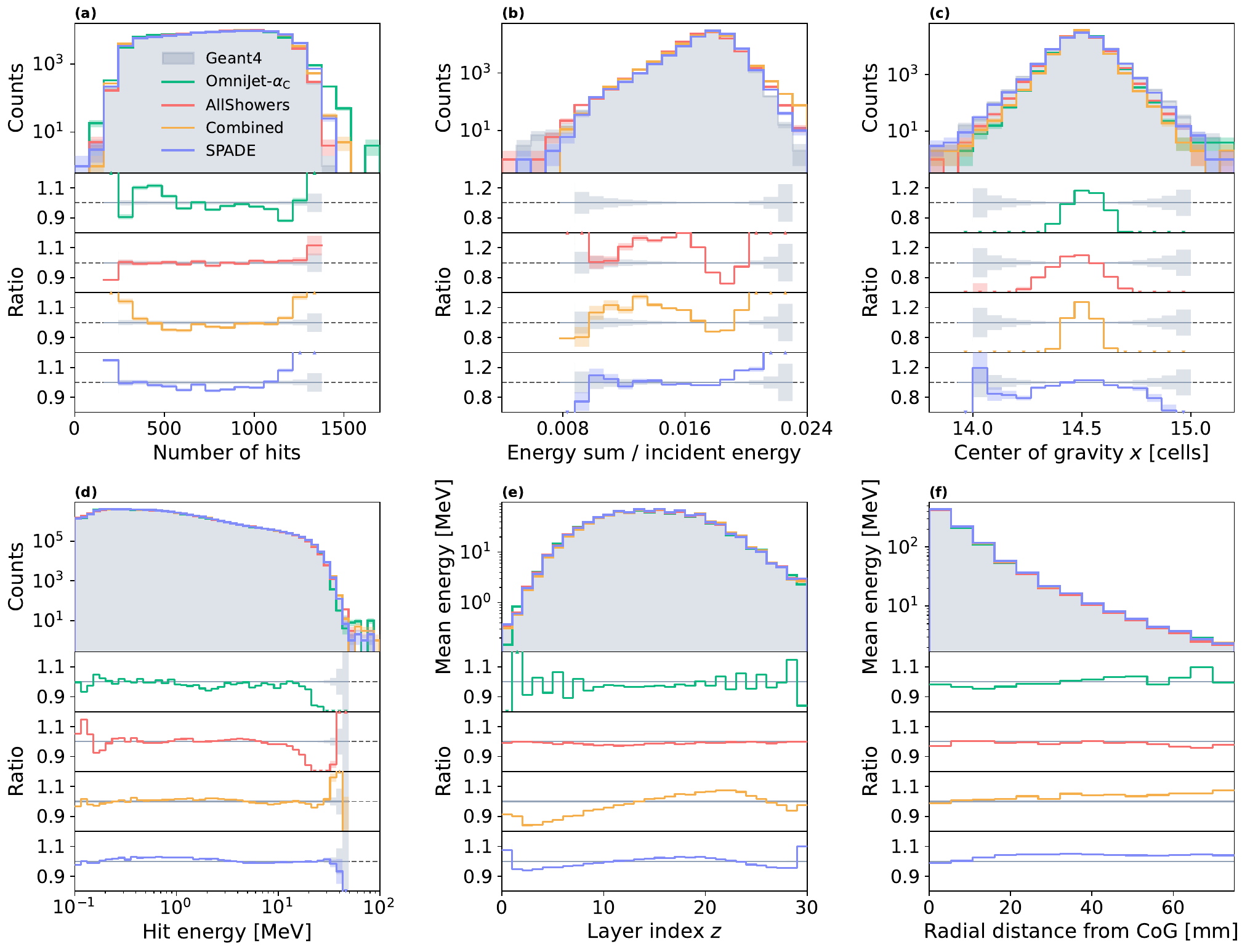}
    \caption{Comparison of \OJAC, \AS, Combined and SPADE on \GH \, photon showers uniformly distributed between 10 and 100 GeV. For each generator, 95k samples are shown. The shaded band represents the statistical standard deviation.}
    \label{fig:GettingHighResults}
\end{figure}

\cref{fig:GettingHighResults} shows the result of training \OJAC, Combined, SPADE and \AS{} on the \GH{} dataset, compared to the Geant4 ground truth. Note that \AS{} predicts the per-layer hit count separately via PointCountFM before generation, giving it a structural advantage on panels~(a) and~(e) that the autoregressive models do not share. 

The top row of \cref{fig:GettingHighResults} shows shower-level observables. For the total number of hits per shower, \cref{fig:GettingHighResults}~(a), \AS{} reproduces the Geant4 distribution most accurately across the full range, largely inheriting this fidelity from its PointCountFM module. In contrast, the three autoregressive models, must determine the sequence length as part of the generation process itself. \OJAC{} samples a dedicated stop token as part of its vocabulary, while Combined and SPADE rely on the stop head described in \cref{subsec:stophead}. \OJAC{} shows the largest deviations, consistent with the fact that the stop decision is entangled with the sampling of physical tokens. Both Combined and SPADE describe the bulk of the distribution well, but deviate more strongly in the tail beyond roughly 1300 hits, suggesting that that the stop head may terminate long showers too late.

The ratio of deposited to incident energy, \cref{fig:GettingHighResults}~(b), is captured best by SPADE, which shows fluctuations only in the tails. Note that this strong agreement relies on the post-processing step described in Appendix~\ref{app:postprocessing}, which actively filters out the $0.35\%$ of generated SPADE showers that suffer from the early-stopping artifact and would otherwise populate an unphysical low-energy tail. Combined and \AS{} fluctuate across the full distribution, including the bulk. \OJAC{} is not conditioned on the incident energy in its current form and is therefore omitted from this panel, although such conditioning could in principle be added.

Spade also best reproduces the center of gravity along $x$, \cref{fig:GettingHighResults}~(c).  Because this metric is an energy-weighted spatial average, it relies heavily on the intra-hit correlation between $x_i$ and $E_i$, features that are separated by two delay steps in the SPADE input encoding. The accurately reproduced distribution therefore validates the design of the staggered conditioning scheme, indicating that the autoregressive structure recovers the energy–position correlation successfully.

The bottom row of \cref{fig:GettingHighResults} shows hit-level observables. The hit-energy spectrum, \cref{fig:GettingHighResults}~(d), shows that SPADE provides the best agreement with Geant4 across three orders of magnitude, followed closely by Combined. Both models predict the hit energy through a dedicated continuous head conditioned on all three spatial coordinates of the same hit, allowing them to model the $E$ spectrum directly. In contrast, \OJAC{} encodes the hit energy jointly with the spatial coordinates through the VQ-VAE codebook, so the achievable energy resolution is bounded by the codebook granularity and visibly degrades in the tails of the spectrum. \AS{} deviates from Geant4 to a similar degree on this observable, which the authors of \AS{} attribute to limited training statistics in the sparsely populated tails of the spectrum~\cite{Buss:2026yrf}.

The mean energy as a function of layer index, \cref{fig:GettingHighResults}~(e), is best reproduced by \AS. This is to be expected, because of the PointCountFM (see \cref{subsec:allshowers}). Among the autoregressive generators, SPADE outperforms both Combined and \OJAC. \OJAC{} in particular shows a pronounced layer-by-layer bumpiness, consistent with the loss of fine-grained spatial information through the VQ-VAE tokenization.

The mean energy as a function of radial distance from the center of gravity, \cref{fig:GettingHighResults}~(f), is best reproduced by \AS{} with some deviation at large radii. SPADE and Combined slightly overestimate the mean hit energy across the full radial range. \OJAC{} displays strong bin-to-bin fluctuations. 
We note that the distribution contains a contribution from the binning of the histogram itself, since the radial distance is a derived quantity computed from discrete sensor positions on the irregular ECAL grid.

\subsection{\GS}
\label{sec:results_gettingsquare}
When evaluating models on \GS{}, each model is trained at the granularity that maximizes its own fidelity at the evaluation resolution. \AS{} is trained at x16 with downstream remapping, Combined and SPADE directly at the evaluation grid (x1-merged). This ensures the comparison reflects each architecture at its best rather than a single imposed protocol; see Appendix~\ref{app:generation_quality} for the supporting evidence and distributions for the remaining granularities. As in the previous subsection, \AS{} serves primarily as an external reference, and the focus here is on whether the factorized SPADE embedding offers an advantage over the joint Combined embedding at the same effective granularity. 

\begin{figure}[t]
    \centering
    \includegraphics[width=1\linewidth]{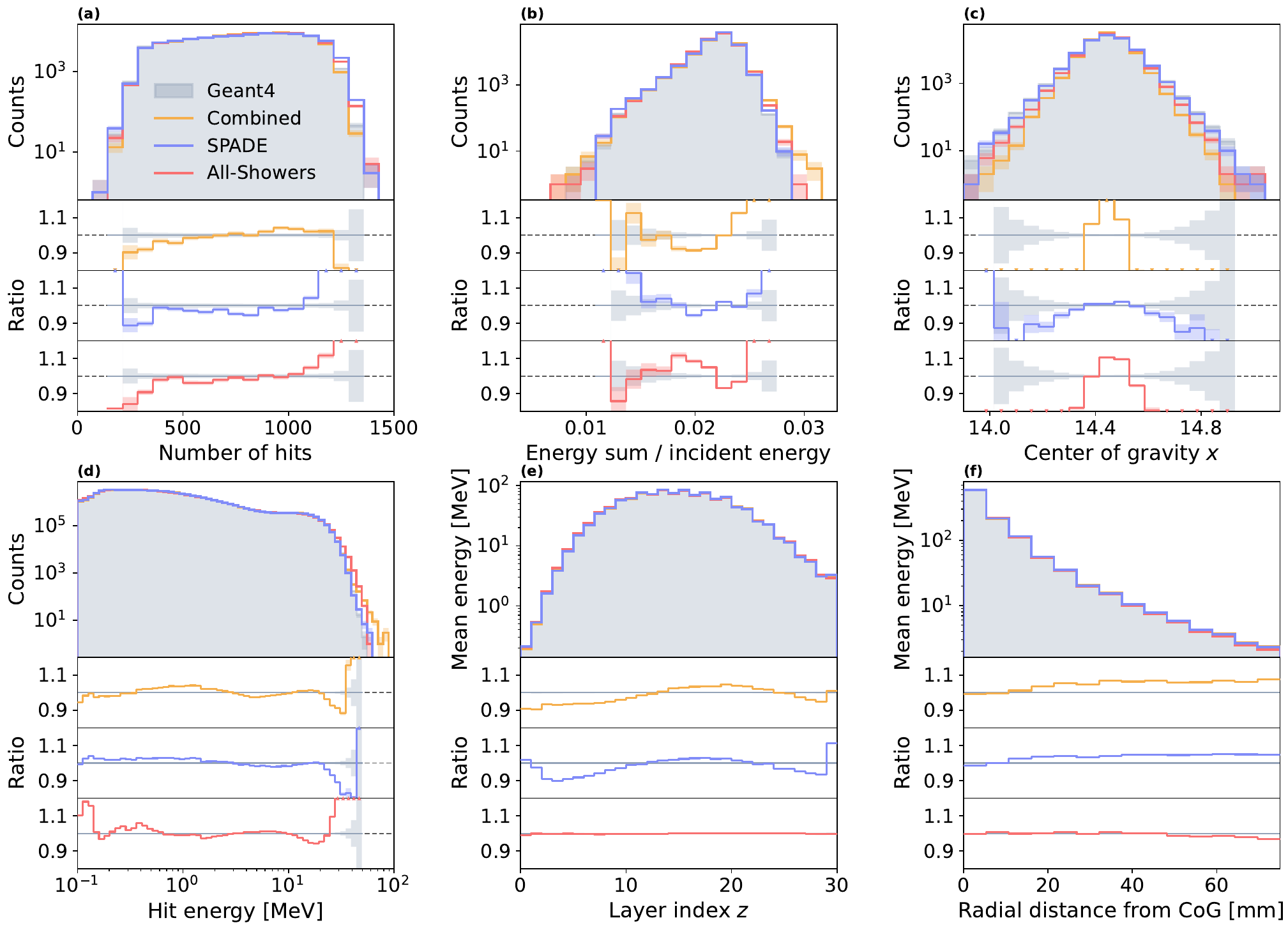}
    \caption{Comparison of \AS, Combined and SPADE on \GS{} photon showers uniformly distributed between 10 and 100 GeV. For each generator, 95k samples are shown. \AS{} was trained on x16 granularity and then remapped to x1-merged, whereas Combined and SPADE were trained on x1-merged directly.}
    \label{fig:GettingSquareResults2}
\end{figure}

The top row of \cref{fig:GettingSquareResults2} shows shower-level observables. For the total number of hits per shower, \cref{fig:GettingSquareResults2}~(a), all three models follow the Geant4 distribution within a few percent across the bulk of the range, with only minor deviations in the high-multiplicity tail. The performance on the ratio of deposited to incident energy, \cref{fig:GettingSquareResults2}~(b), is similar for all three models, with comparable levels of fluctuation around the truth. SPADE is arguably best in the bulk of the distribution.

The center of gravity along $x$, \cref{fig:GettingSquareResults2}~(c), shows a clear separation between the two autoregressive models: SPADE follows the Geant4 distribution closely across the central peak, while Combined exhibits noticeably larger deviations in the ratio panel. As already discussed for \GH, this observable is an energy-weighted spatial average and therefore directly sensitive to the intra-hit $(x, E)$ correlation. The SPADE advantage here is consistent with the factorized embedding distributing the training signal more uniformly across the spatial vocabulary: with $N_x + N_y + N_z$ rather than $N_x \cdot N_y \cdot N_z$ unique spatial vectors, every embedding is updated frequently, and the energy–position correlation is recovered without the token sparsity that affects Combined. \AS{} produces a slightly narrower distribution than Geant4, visible as an excess at the peak and a deficit in the tails of the ratio panel; SPADE matches Geant4 more closely than \AS{} across the full region.

The bottom row of \cref{fig:GettingSquareResults2} shows hit-level observables. The hit-energy spectrum, \cref{fig:GettingSquareResults2}~(d), is reproduced with comparable fidelity by all three models across nearly three orders of magnitude, with only minor disagreements in the sparsely populated high-energy tail. The mean energy as a function of layer index, \cref{fig:GettingSquareResults2}~(e), is most accurately reproduced by \AS{} due to its PointCountFM module, while SPADE and Combined behave essentially equivalently, both showing a mild dip relative to Geant4 in the middle layers. For the mean energy as a function of radial distance from the center of gravity, \cref{fig:GettingSquareResults2}~(f), \AS{} reproduces the Geant4 distribution well across the full radial range. Combined develops a systematic upward drift toward large radii, overshooting Geant4 by up to roughly 10\% at the outermost bins, while SPADE shows a slightly more moderate version of the same trend.

\section{Training efficiency}
\label{sec:training_efficiency}
\begin{figure}[t]
    \centering
    \includegraphics[width=.99\linewidth]{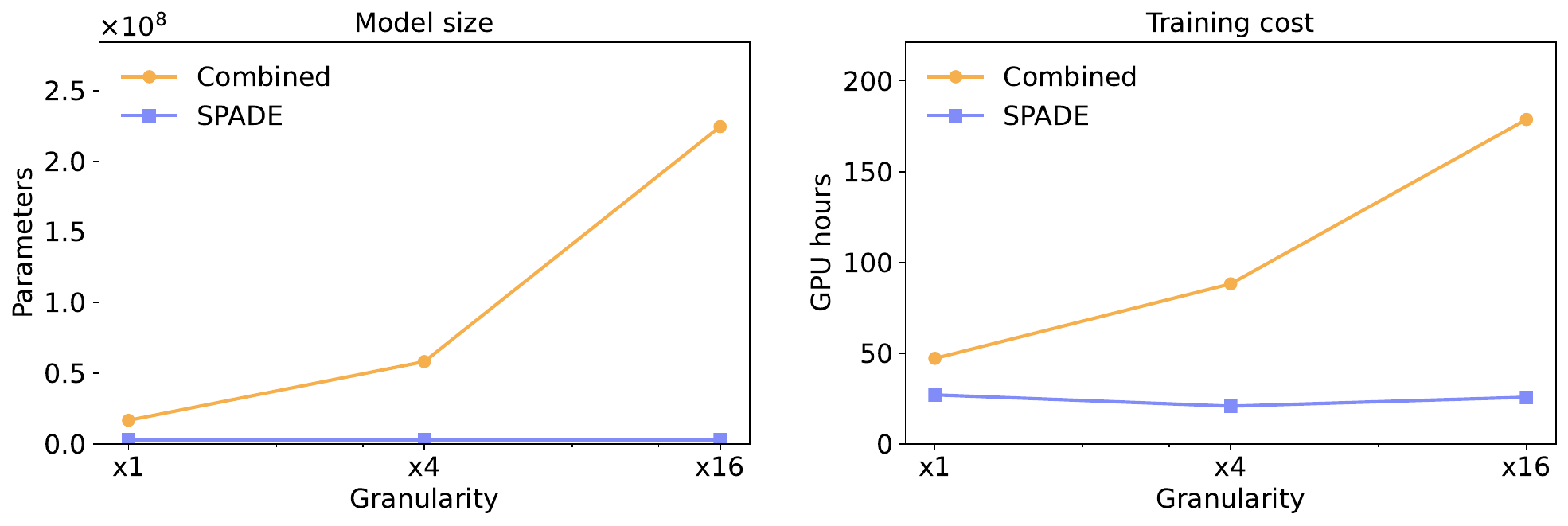}
    \caption{Left: the number of model parameters as a function of the granularity for SPADE and Combined. Right: the number of GPU hours needed to train each model within 2\% of the the minimal loss reached after 55k steps.}
    \label{fig:scaling}
\end{figure}

Due to the difference in tokenization strategy, the Combined model is substantially larger than SPADE. As shown in \cref{sec:combined}, Combined's spatial embedding layer and token-prediction head scale as $N^3$ with the granularity $N$, while SPADE's factorized counterparts scale as $3N$. The resulting parameter counts at each granularity are shown in \cref{fig:scaling} (left) and tabulated in \cref{tab:scaling}, which additionally lists the granularity-independent transformer backbone shared by both models.

To compare wall-clock training cost on a fair basis across granularities and architectures, we define it as the number of GPU hours (NVIDIA H100) needed for each model to first reach within 2\% of its own minimum validation loss within 55k training steps. The threshold is per-model rather than absolute, because SPADE and Combined converge to noticeably different minima at finer granularities. The 2\% level corresponds to essentially-converged training, beyond which further steps yield diminishing returns. The resulting cost per model is shown in \cref{fig:scaling} (right) and tabulated in \cref{tab:scaling}: at \GS-x16, Combined requires 178.7~GPU\,hours against SPADE's 25.8~GPU\,hours, a factor of 6.9 more.

Two effects compound to produce this gap. First, Combined's large parameter count limits the number of training showers that fit on a single GPU, forcing the use of batch accumulation to reach the desired effective batch size. These extra accumulation steps translate directly into wall-clock cost. Second, SPADE needs fewer gradient steps to converge, a representational advantage of the factorized embedding. By dividing
the output space across three smaller vocabularies instead of one large
one, SPADE mitigates gradient sparsity and updates every token embedding
more frequently.

The validation-loss trajectories underlying these numbers, including the markers at which each run first crosses the 2\% threshold, are shown in  Appendix~\ref{app:convergence} . The same curves also reveal that SPADE converges to a lower minimum validation loss than Combined at finer granularities.
\vspace{-2mm}
\begin{table}[h]
\caption{Model size and training cost vs.\ granularity. Combined's and SPADE's transformer block is granularity-independent; the remaining parameters come from granularity-dependent input/output heads At x16, SPADE uses a factor of 74 fewer parameters and trains 6.9 times faster than Combined.}
\centering
\begin{tabular}{lcccccc}
 & \multicolumn{3}{c}{Parameters} & \multicolumn{3}{c}{GPU hours} \\
\cmidrule(lr){2-4}\cmidrule(lr){5-7}
Model & x1 & x4 & x16 & x1 & x4 & x16 \\
\midrule
Combined & 16.80\,M & 58.35\,M & 224.56\,M & 47.2 & 88.2 & 178.7 \\
SPADE    & 2.96\,M  & 2.98\,M  & 3.02\,M   & 27.1 & 20.9 & 25.8 \\
\quad of which: transformer blocks & 2.81\,M & 2.81\,M & 2.81\,M & --- & --- & --- \\
\bottomrule
\end{tabular}
\label{tab:scaling}
\end{table}
\vspace{-2mm}

\section{Inference time}
\label{sec:inference_time}
We measure the wall-clock time to generate a single shower for Combined, SPADE, and \AS{} on a single NVIDIA A100-SXM4-80GB, sweeping incident energies (10, 55, 100~GeV) at batch size 4096 --- the largest batch size that fits on a single GPU for all three models. We discard one warmup batch per setting and average over $N = 10$ independent batches, reporting mean $\pm$ standard error of the mean. The reported per-shower time includes autoregressive token generation and the mapping of tokens onto the \GS-x1-merged geometry. To ensure a comparison at their respective highest fidelities, \AS{} is generating showers in the \GS{}-x16 geometry, while SPADE and Combined are generating coarser showers in the \GS{}-x1-merged representation. The postprocessing step described in Appendix~\ref{app:postprocessing} is not included in the inference time for SPADE, as it only effects $0.35\% $ of showers. The results are summarized in \cref{tab:timing}.

\AS{} is consistently the fastest model, generating showers roughly three times faster than SPADE and up to five times faster than Combined. This speedup occurs because \AS{} is not an autoregressive model generating individual hits sequentially; instead, it iterates over a small, fixed number of flow-matching integration steps. However, its generation time still grows with incident energy, because the computational cost of each network pass increases with the overall size of the generated point cloud.

In contrast, SPADE and Combined perform one cached transformer step per sequence position, with the KV cache growing linearly with the number of generated hits. Their per-shower time therefore scales directly with the average hit count, driving the increase in generation time at higher incident energies. Between the two autoregressive models, SPADE generates roughly twice as fast as Combined at both 55 and 100~GeV. This is consistent with the parameter scaling detailed in \cref{tab:scaling}: Combined's much larger embedding and output heads introduce a per-step cost that compounds over the long autoregressive sequences.
\vspace{-2mm}
\begin{table}[h]
\centering
\caption{Per-shower inference time in ms/shower on a single NVIDIA A100-SXM4-80GB at batch size 4096, as a function of incident energy and model. Values are mean $\pm$ SEM in ms/shower across $N=10$ independent batches.}
\label{tab:timing}
\begin{tabular}{c r@{\,$\pm$\,}l r@{\,$\pm$\,}l r@{\,$\pm$\,}l}
\toprule
$E_\mathrm{inc}$ [GeV] & \multicolumn{2}{c}{SPADE} & \multicolumn{2}{c}{Combined} & \multicolumn{2}{c}{\AS{}} \\
\midrule
10  & 12.72 & 0.05 & 15.60 & 0.16 & 3.71 & 0.05 \\
55  & 21.71 & 0.04 & 46.15 & 0.17 & 7.06 & 0.66 \\
100 & 33.44 & 0.12 & 62.99 & 0.01 & 11.90 & 0.01 \\
\bottomrule
\end{tabular}
\end{table}
\vspace{-2mm}
\section{Conclusion}
\label{sec:conclusions}

As autoregressive transformer models are increasingly being used outside of the language domain for which they were originally developed, it is important to revisit the original architectural strategies. Tokenization has been the go-to method to turn text into numbers that the model can ingest, but in domains like particle physics, where our input already is numerical, this might not be the ideal choice. Furthermore, physics data often lives in high-dimensional spaces, parts of which can be discrete. In this work we have explored how to deal with such a high-dimensional space without being bottlenecked by giant codebooks, as well as how to preserve the important correlations between the different features. 

SPADE is applied to the crucial problem of simulating highly granular calorimeter data within a feasible compute budget. The introduction of the split head approach of SPADE enables the model to directly make use of the discrete features as-is. Compared to the original \OJAC{} model, it completely eliminates the need to train a VQ-VAE for tokenization. The feature correlations are preserved through the delay mechanism, where the different features become conditioned on one another. In comparison to the Combined model, which combines the spatial coordinates into a single number, SPADE enables easy scaling to higher granularities. In contrast, a higher granularity for the Combined model leads to a drastic inflation of the number of model parameters, which can make training and inference prohibitively expensive. We have shown that %
SPADE is easier and more efficient to train than both \OJAC{} and Combined, 
and that these technical improvements introduced here allow an autoregressive model for the first time to achieve a performance on par with state of the art models in generating realistic electromagnetic showers, without the penalty of a large Combined codebook. 

The task of additionally simulating hadronic showers, which is a challenge made harder by the higher complexity of the data, is left for future work. Another avenue for future work relates to the inference time. Despite improving compared to the Combined model, SPADE is still slower than state of the art models like \AS{} due to the inherent sequential nature of autoregressive generation. However, speed is not the main point of SPADE. Rather it is the improvements over single-stream combined tokenization models, which, given the increased importance of autoregressive transformers for foundation models, is highly significant for models across a wide range of domains.

Beyond calorimeter simulation, the SPADE mechanism (split-and-delay embeddings) applies to any sequential generative task whose tokens carry multiple features, whether discrete or continuous. By embedding each feature independently and recovering their correlations through staggered conditioning, SPADE extends the next token prediction paradigm of large language models to multi-dimensional data, without lossy intermediate representations. This positions SPADE as a candidate building block for LLM-style pretraining-and-fine-tuning workflows in HEP and beyond, in which a single autoregressive backbone, pretrained on large unlabeled datasets of multi-feature tokens, is subsequently adapted to new detectors, particle species, or different problem domains with multi-dimensional sensor data. An immediate application could be in high-dimensional astrophysics data, where for example data from the Square-Kilometer-Array has already been used to train autoregressive transformer models~\cite{Heneka:2025fpe}.

\newpage

\appendix
\section{Model details and hyperparameters}
\label{sec:appendix_hyperparams}
Model hyperparameters can be found in \cref{tab:hparams_models}. The standardization statistics $\mu$ and $\sigma$ are computed once on the training set and applied to validation and test data unchanged.
\begin{table}[h]
\centering
\caption{Hyperparameters for the SPADE and Combined models. The two models differ only in their embedding and unembedding scheme (top), while the transformer backbone, training, and sampling configuration are identical (bottom).}
\label{tab:hparams_models}

\begin{tabular}{llcc}
\toprule
Type & Parameter & SPADE & Combined \\
\midrule
\multirow{6}{*}{Embedding / Unembedding}
  & input representation               & per-axis $(x,y,z)$ & joint voxel index \\
  & vocabulary size\,$^\dagger$        & $N_x + N_y + N_z$ & $N_x N_y N_z$ \\
  & \quad at \GH{}/\GS-x1    & 96                 & 27\,002 \\
  & \quad at \GS-x4                    & 156                & 108\,002 \\
  & \quad at \GS-x16                   & 276                & 432\,002 \\
  & output heads                       & three softmaxes    & single softmax \\
\bottomrule
\end{tabular}
\vspace{0.5em}
\begin{minipage}{\linewidth}
\footnotesize
$^\dagger$ The base vocabulary size is determined by the granularity. An additional two tokens (+2) account for the padding and missing-ID embeddings (see \cref{eq:combined_token,eq:spade_token}). These are added once globally for the Combined model, and per axis for SPADE (resulting in +6).
\end{minipage}
\vspace{1em}

\begin{tabular}{lll}
\toprule
Type & Parameter & Value \\
\midrule
\multirow{3}{*}{Data Preprocessing}
  & hit energy      & $\log E$, then $(x - \mu)/\sigma$ \\
  & incident energy & $(x - \mu)/\sigma$ \\
  & hit ordering    & layer-ascending, energy-descending \\
\midrule
\multirow{8}{*}{Backbone}
  & embedding dim         & 256 \\
  & layers                & 3 \\
  & attention heads       & 8 \\
  & MLP ratio             & 4 \\
  & positional encoding   & RoPE \\
  & attention             & multi-query, FlashAttention \\
  & dropout               & 0 \\
  & max sequence length   & 1700 for \GH{} and 4600 for \GS{}\\
\midrule
\multirow{3}{*}{Energy head (MoG)}
  & mixture components $K$ & 4 \\
  & MLP layers             & 2 \\
  & MLP hidden dim         & 128 \\
  & relative energy noise $\sigma_{\mathrm{noise}}$ & 0.05 \\ 
\midrule
\multirow{6}{*}{Stop head}
  & spatial embedding dim  & 16 \\
  & energy projection dim  & 16 \\
  & position projection dim & 16 \\
  & hidden layers          & 2 \\
  & hidden dim             & 128 \\
  & dropout                & 0.2 \\
  & BCE positive class weight   & 1000 \\
\midrule
\multirow{8}{*}{Training}
  & optimizer            & Ranger (Lookahead + RAdam) \\
  & Adam $\beta_1, \beta_2$ & 0.95, 0.999 \\
  & Lookahead $\alpha, k$ & 0.25, 16 \\
  & learning rate        & $1 \times 10^{-4}$ (constant after warmup) \\
  & warmup steps         & $2\,000$ \\
  & weight decay         & $5 \times 10^{-3}$ \\
  & per-GPU batch size   & 48 \\
  & gradient clipping    & 1.0 \\
\midrule
\multirow{2}{*}{Inference}
  & weights              & EMA, decay 0.999 \\
  & sampling temperature & 1.0 \\
  & stop threshold & $p > 0.98$ \\
\bottomrule
\end{tabular}
\end{table}

\section{Postprocessing}
\label{app:postprocessing}
Autoregressive models can accumulate errors step by step, and once the sampled shower drifts too far from the training distribution, the model can \emph{hallucinate} showers that no longer resemble physically valid ones. In our setting the dominant failure mode is the total deposited energy $E_\mathrm{sum} = \sum_i E_i$ falling far below the range observed in Geant4 at the same incident energy. To mitigate this we fit a lower envelope to the Geant4 showers $E_\mathrm{sum}$ vs.\ $E_\mathrm{inc}$ distribution and reject generated samples that fall below it.
A shower is flagged if:
\begin{equation}
  E_\mathrm{sum} \;<\; s \cdot \left( a\, E_\mathrm{inc}^{\,b} + c \right),
  \label{eq:energy_sum_filter}
\end{equation}
where $(a, b, c)$ are fit to a low percentile of the $E_\mathrm{sum}$ vs.\ $E_\mathrm{inc}$ relation in Geant4, and the safety factor $s = 0.9$ pushes the threshold further below the fitted envelope. For the \GS{} datasets we use $a = 0.0174$, $b = 0.9379$, $c = -0.0180$ (with $E_\mathrm{inc}$ and $E_\mathrm{sum}$ in GeV), fit to the lower envelope of the Geant4 $E_\mathrm{sum}$--$E_\mathrm{inc}$ relation. Flagged showers are regenerated under the same conditioning and replace the original in place; we allow up to $k_\mathrm{max} = 2$ regeneration rounds, after which any still-flagged shower is kept unchanged. Applying the same threshold to the \GS{} test set flags $f_\mathrm{G4} = 0.014\%$ of real showers, with the rate roughly flat across incident energy bins (between $0.006\%$ and $0.026\%$ across the 10--90~GeV range).

On generated \GS{} showers, the first-pass flag rate is $f_\text{SPADE} = 0.35\% \pm 0.03\%$ (Poisson on $\sim$140/40,000 counts) for SPADE and $ f_\text{Combined} = 0.015\% \pm 0.006\%$ (Poisson on $\sim$6/40,000 counts) for Combined, both falling to $0\%$ after $k_\mathrm{max} = 2$ regeneration rounds. We treat the residual bias as a known systematic of the procedure. Combined's flag rate is statistically indistinguishable from the Geant4 baseline, indicating that this failure mode is essentially absent for Combined and specific to SPADE. The cause of this difference between the two encodings is not yet understood, but the postprocessing filter removes the artifact for SPADE.

\section{Generation quality across granularities}
\label{app:generation_quality}
\begin{figure}[p]
    \centering
    \includegraphics[width=0.85\linewidth]{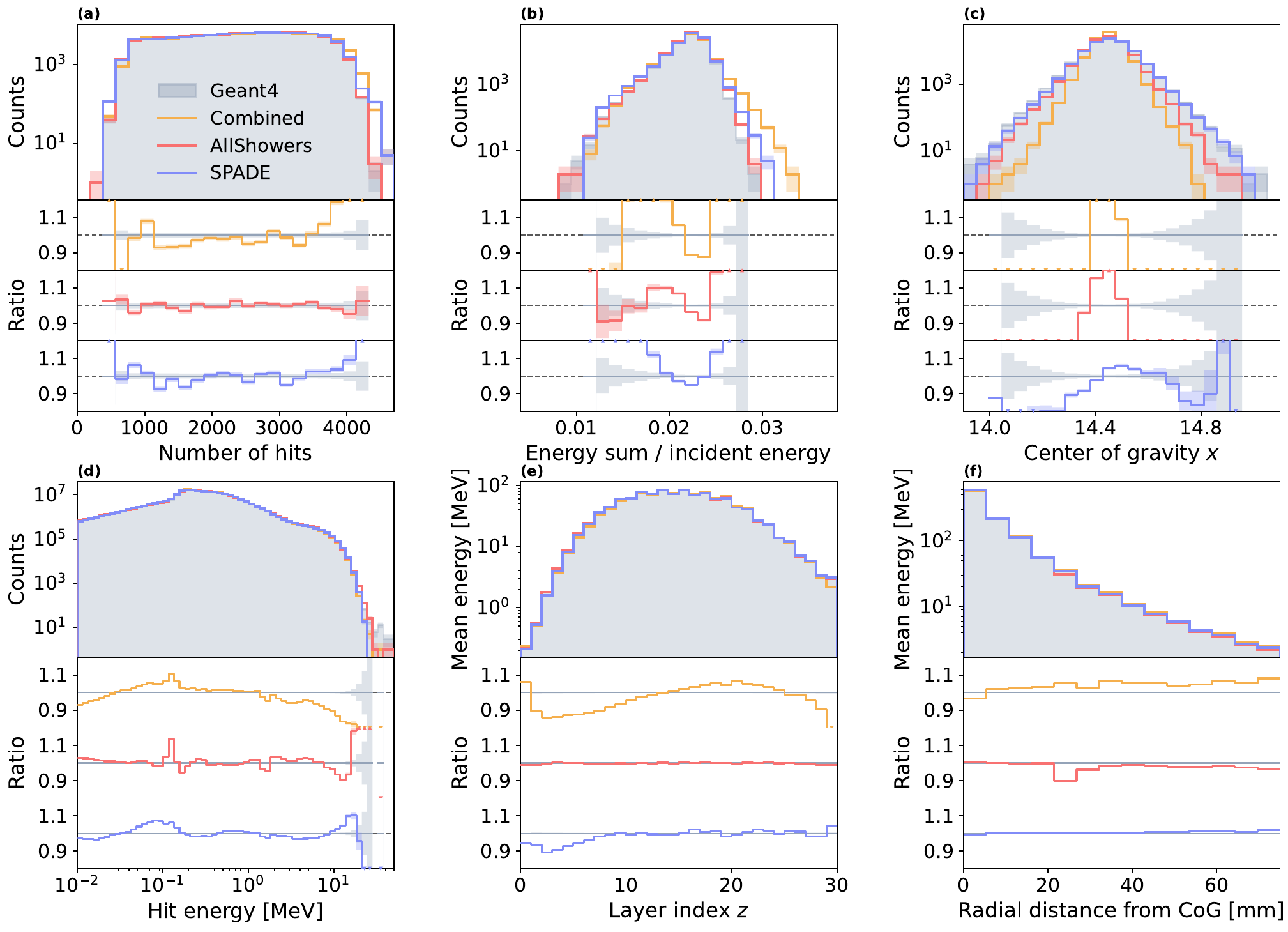}
    \caption{Trained on \GS-x1 at native resolution ($30\times30\times30$). Comparison of \AS, Combined and SPADE on photon showers uniformly distributed between 10 and 100~GeV. For each generator, 95k samples are shown; the shaded band is the statistical standard deviation.}
    \label{fig:x1}
\end{figure}

\begin{figure}[p]
    \centering
    \includegraphics[width=0.85\linewidth]{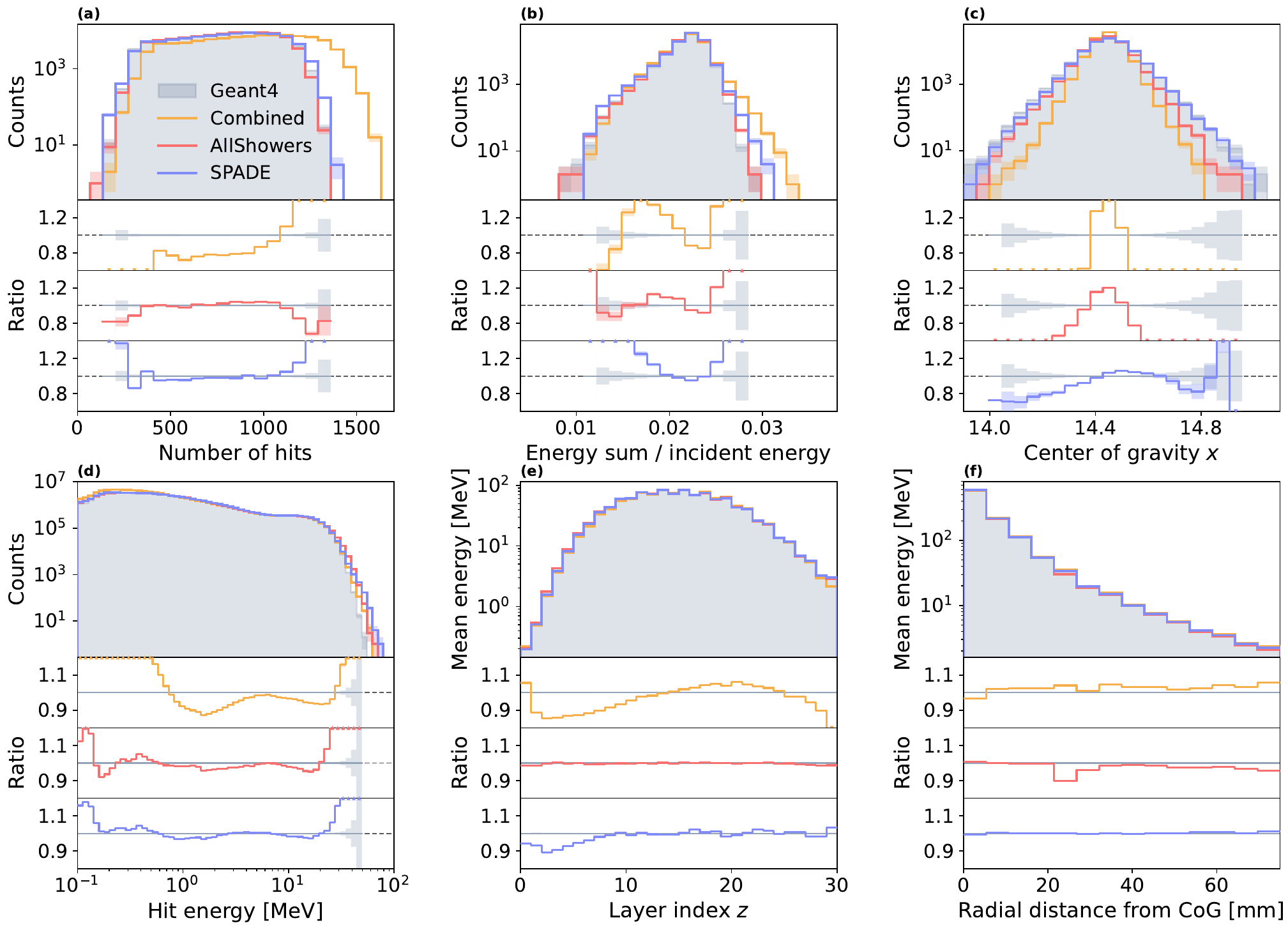}
    \caption{Models trained on \GS-x1 with generated showers merged to the x1-merged grid. Otherwise as \cref{fig:x1}.}
    \label{fig:x1_remapped}
\end{figure}

\begin{figure}[p]
    \centering
    \includegraphics[width=0.85\linewidth]{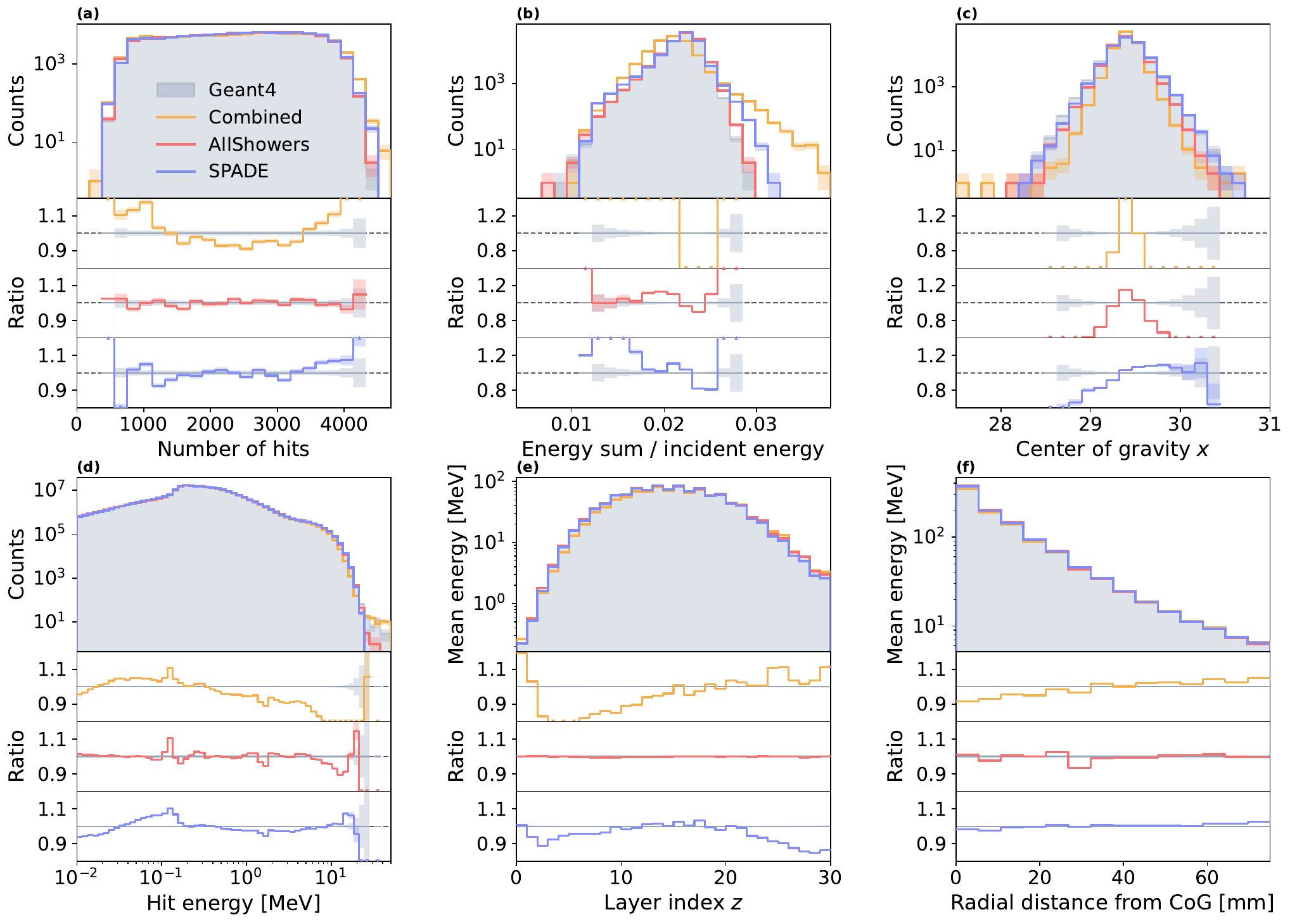}
    \caption{Trained on \GS-x4 at native resolution ($60\times60\times30$). Comparison of \AS, Combined and SPADE on photon showers uniformly distributed between 10 and 100~GeV. For each generator, 95k samples are shown; the shaded band is the statistical standard deviation.}
    \label{fig:x4}
\end{figure}

\begin{figure}[p]
    \centering
    \includegraphics[width=0.85\linewidth]{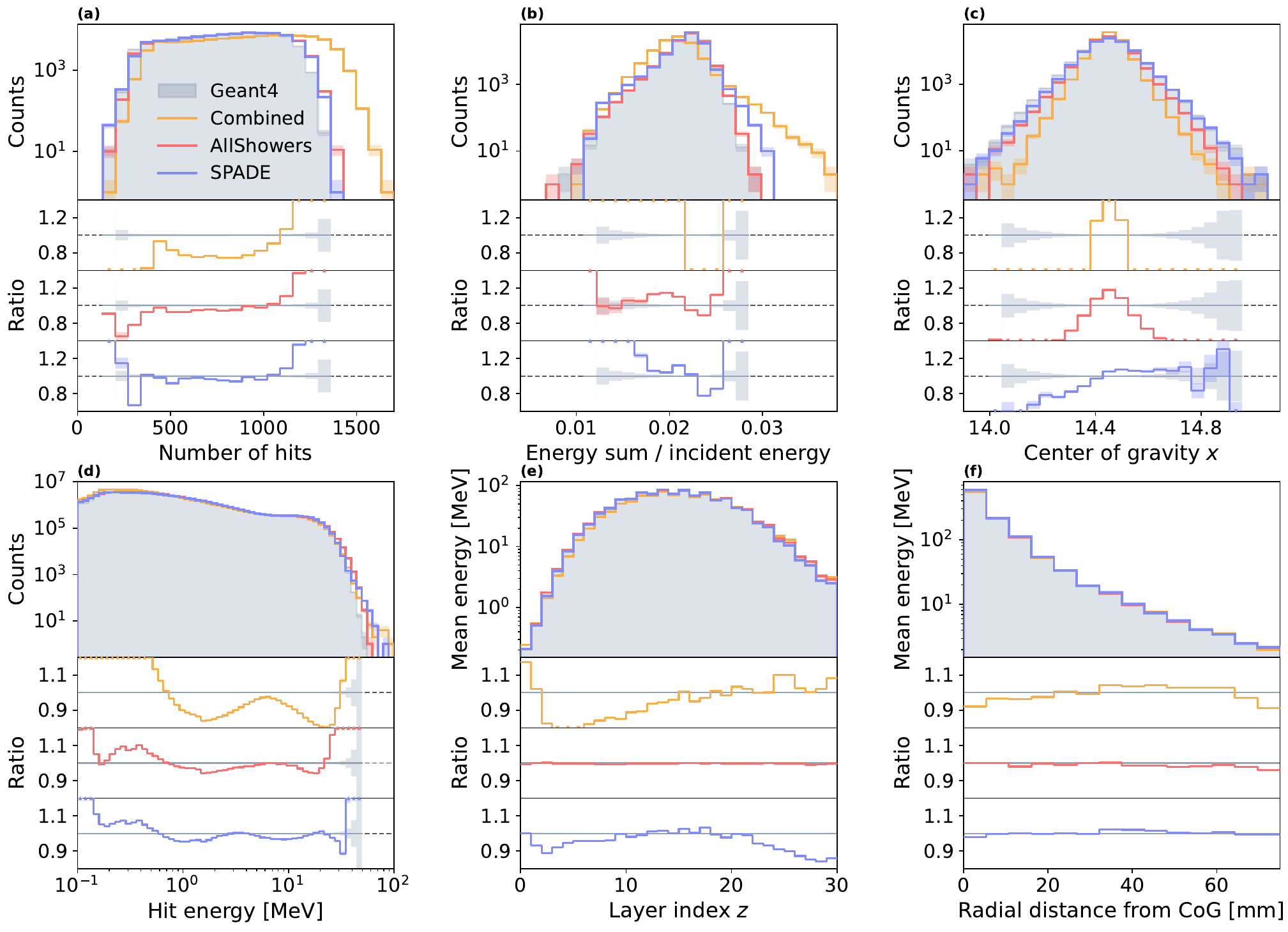}
    \caption{Models trained on \GS-x4 with generated showers remapped to the x1-merged grid. Otherwise as \cref{fig:x4}.}
    \label{fig:x4_remapped}
\end{figure}

\begin{figure}[p]
    \centering
    \includegraphics[width=0.85\linewidth]{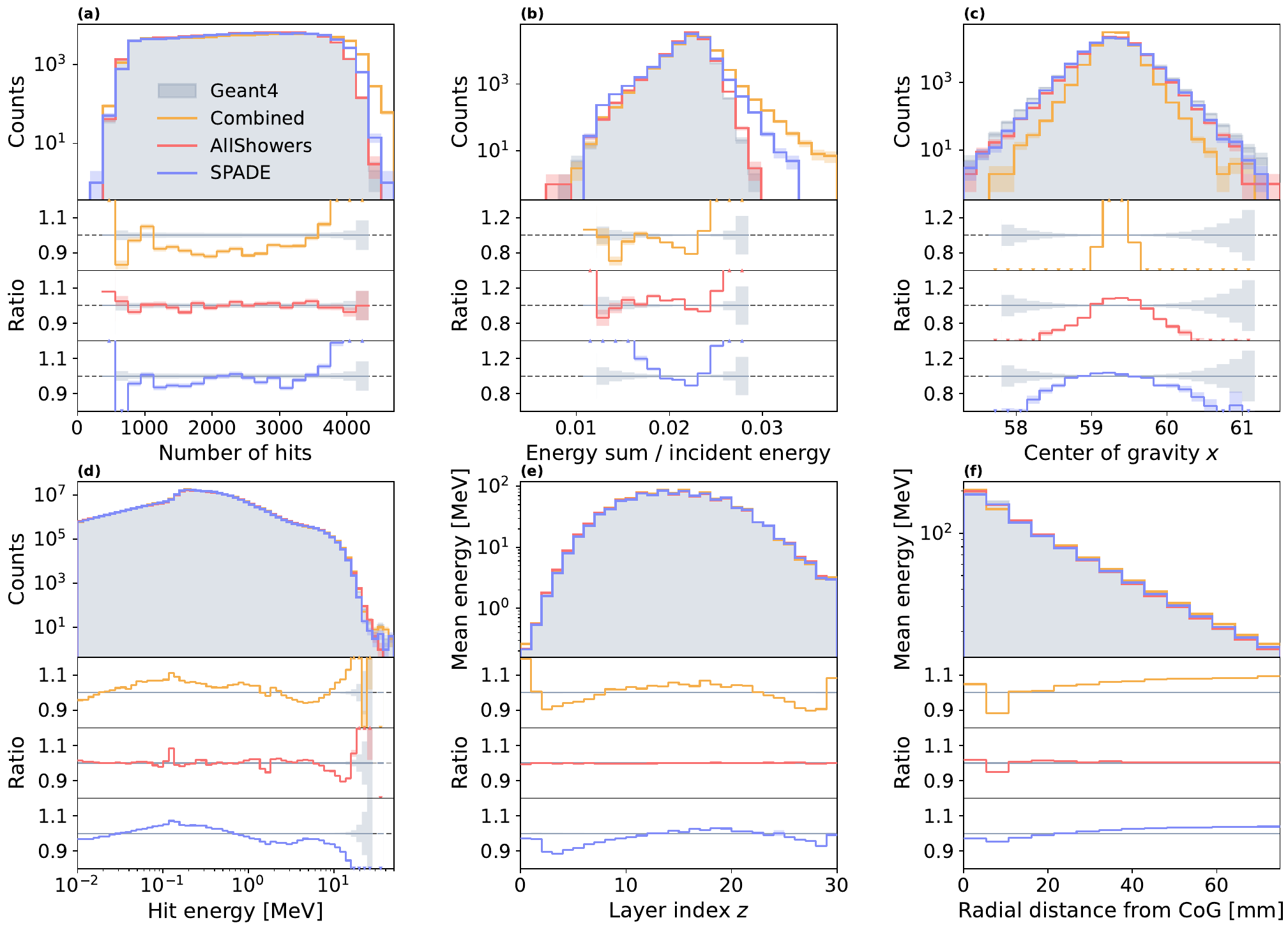}
    \caption{Trained on \GS-x16 at native resolution ($120\times120\times30$). Comparison of \AS, Combined and SPADE on photon showers uniformly distributed between 10 and 100~GeV. For each generator, 95k samples are shown; the shaded band is the statistical standard deviation.}
    \label{fig:x16}
\end{figure}

\begin{figure}[p]
    \centering
    \includegraphics[width=0.85\linewidth]{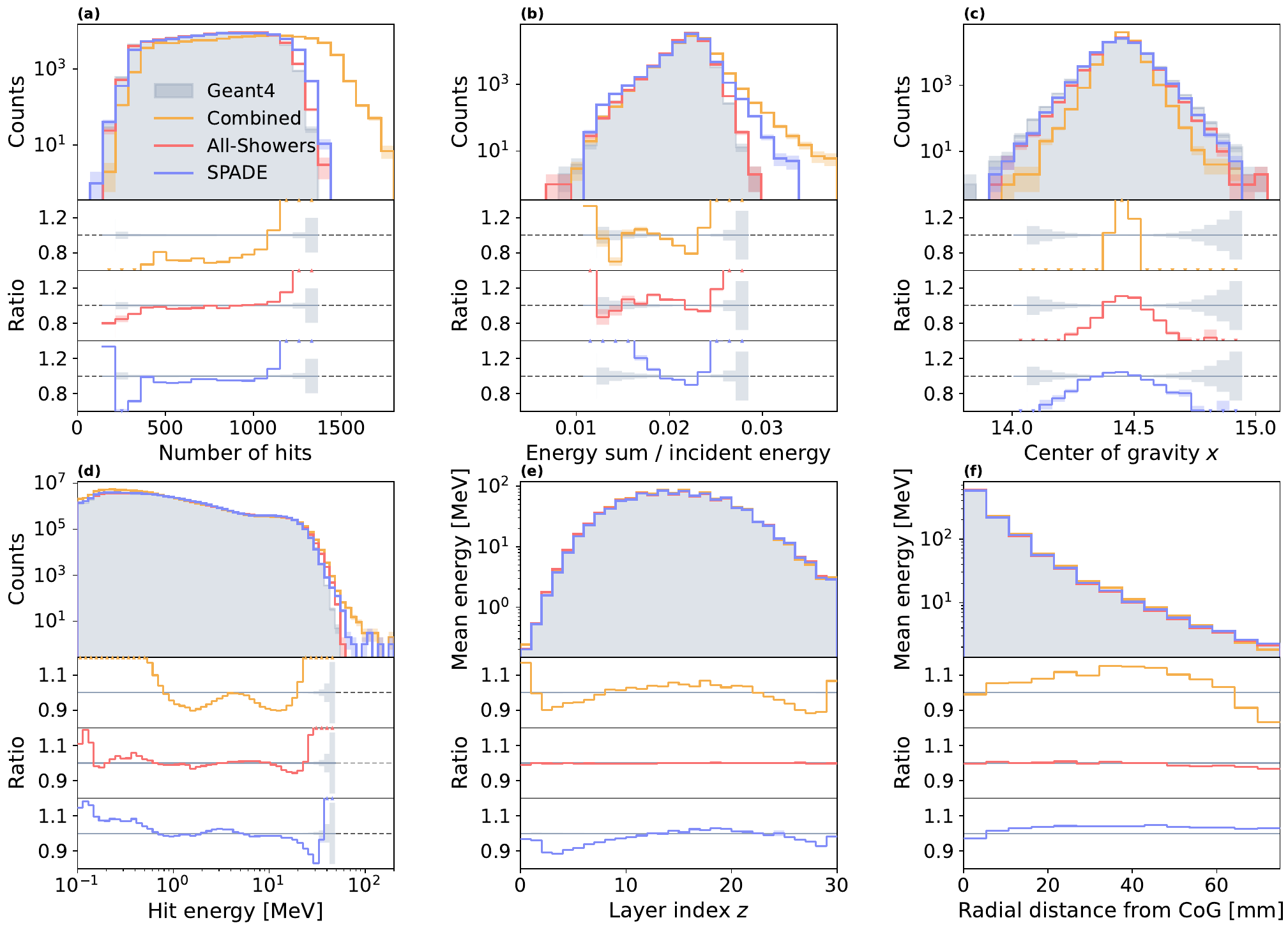}
    \caption{Models trained on \GS-x16 with generated showers remapped to the x1-merged grid. Otherwise as \cref{fig:x16}.}
    \label{fig:x16_remapped}
\end{figure}

\cref{sec:results_gettingsquare} focused on the \GS-x1-merged representation, the common grid on which all three models can be compared directly. Here we provide the full set of distributions for the remaining \GS{} variants. For each transverse granularity we show two figures: the models trained and evaluated at their \emph{native} resolution (\cref{fig:x1,fig:x4,fig:x16}), and the same generated showers after a post-sampling remapping onto the x1-merged grid (\cref{fig:x1_remapped,fig:x4_remapped,fig:x16_remapped}). For the x4 and x16 granularities, the remapping re-bins the higher-resolution output down to $30\times30\times30$ and merges hits that share a voxel. For x1, the grid is unchanged and only the merging step applies.

The central observation across these six configurations is that Combined and SPADE retain their performance as the native granularity increases: at x1, x4 and x16 (\cref{fig:x1,fig:x4,fig:x16}) both models reproduce the Geant4 reference well. Remapping a model trained at a higher granularity down to the x1-merged grid, however, does not improve the agreement, and the two encodings respond to it very differently. SPADE stays close to the Geant4 reference across the remapped configurations, but Combined degrades visibly, most clearly in the hit multiplicity (\cref{fig:x1_remapped}(a),~\cref{fig:x4_remapped}(a) and~\cref{fig:x16_remapped}(a)): merging onto a coarser voxel grid barely changes its hit count, meaning too few of its generated hits coincide on the same voxel to be collapsed by the merge. Combined therefore samples spatial positions too diffusely across the vocabulary, missing the dense local clusters that characterize real showers. The factorized SPADE embedding, by contrast, learns these spatial correlations sharply enough that its merged hit count matches Geant4 — evidence that the factorization helps SPADE capture the actual hit geometry, not merely the aggregate distributions.

These remapping results also explain a methodological choice in the main comparison. In \cref{sec:results_gettingsquare} each model is evaluated under the training strategy that maximizes its own fidelity, so that the comparison reflects each architecture at its best rather than a single imposed protocol. The present results show what that strategy is for the autoregressive models: training at a higher granularity and remapping down brings no gain for SPADE and breaks the hit multiplicity for Combined, so both are trained directly on x1-merged. \AS{} sits at the opposite end of this trade-off. As noted in \cref{sec:results_gettingsquare}, it is trained at x16 and remapped to x1-merged, because training it directly at the lowest granularity produces artifacts (\cref{fig:x1}(f) and~\cref{fig:x1_remapped}(f)). This is consistent with the larger literature, where diffusion and flow-matching-based generators are generally more accurate when trained at the highest available granularity and mapped back to the detector space~\cite{Buhmann_2023}. SPADE is the exception: it is most accurate trained directly at the evaluation granularity, with no high-resolution intermediate stage at all.

One caveat remains. Of the native datasets used here, only x16 is free of duplicate hits sharing a voxel; the x1 and x4 variants retain duplicates that no physically realizable readout could record. A real detector at those granularities would therefore measure a lower hit multiplicity per shower than the corresponding training data. The remapped variants collapse these duplicates and are physical in this sense. Whether the granularity behavior reported here generalizes to other detector geometries remains to be evaluated. Even so, the comparison provides a useful first characterization of how the factorized (SPADE) and joint (Combined) encodings respond to changes in detector segmentation.



\newpage
\section{Convergence}
\label{app:convergence}
\begin{figure}[t]
    \centering
    \includegraphics[width=1\linewidth]{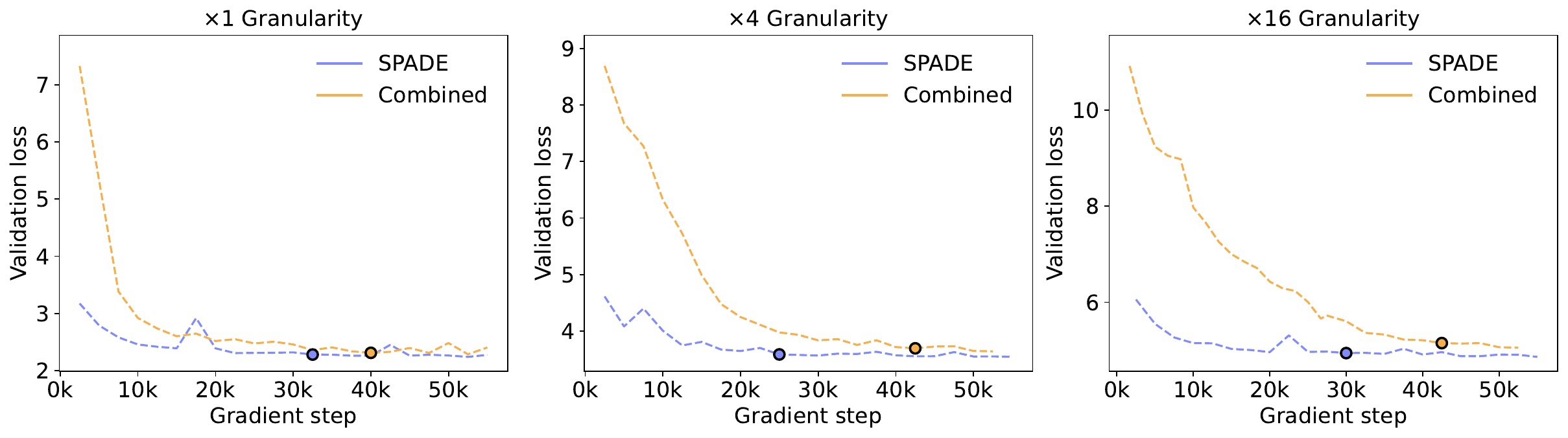}
    \includegraphics[width=1\linewidth]{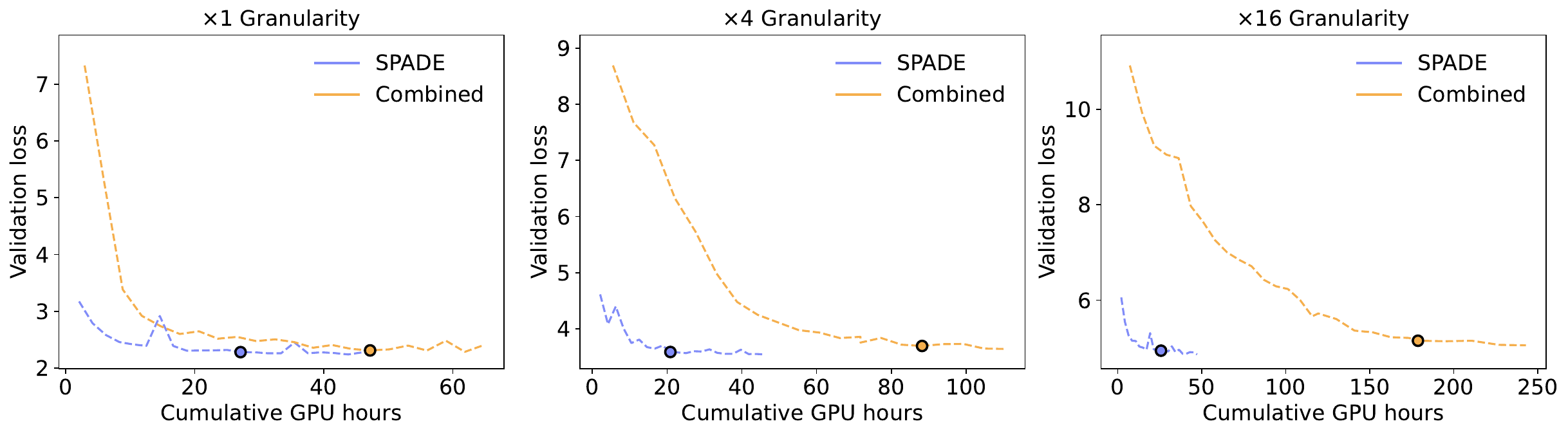}
    \caption{Validation loss for SPADE and Combined at the three \GS{} granularities, versus gradient step (top) and cumulative GPU hours on an NVIDIA H100 (bottom). Markers: first point at which each run reaches within 2\% of its own minimum. SPADE matches or undercuts Combined's minimum validation loss and reaches it at a much lower GPU-hour cost at every granularity, with the gap widening with finer segmentation.}
    \label{fig:convergence}
\end{figure}

\cref{fig:convergence} shows the validation-loss trajectories underlying the training-efficiency summary of \cref{sec:training_efficiency}, plotted against the gradient step (top) and cumulative GPU hours (bottom). The markers indicate the first point at which each run reaches within 2\% of its own minimum, the threshold used in \cref{tab:scaling} and \cref{fig:scaling}.

The trajectories make two features visible that the aggregate numbers do not capture. First, SPADE descends more steeply than Combined from the earliest steps. This is the visual signature of the reduced gradient sparsity discussed in \cref{sec:training_efficiency}. Second, the gap between the two dotted reference lines widens with granularity: at x1 the two minima are essentially tied, while at x16 SPADE settles a clear margin below Combined. The factorized embedding therefore yields not only a faster route to convergence but also a lower endpoint at higher detector granularity.

\section*{Code Availability}
The code for this paper can be found at \url{https://github.com/uhh-pd-ml/SPADE}.
\section*{Data Availability}
The code for recreating the dataset can be found at \url{https://github.com/FLC-QU-hep/getting_high}. 

\ack{We thank Sarah Heim for helpful comments on the manuscript. We thank Thorsten Buss for valuable discussions on FlashAttention and training strategies, and for his help in providing additional scripts to adapt the AllShowers model to the new datasets. We thank Anatolii Korol for providing the data used to create the GettingSquare dataset.}

\funding{J.B., A.H., G.K., M.M. and H.R. are supported by the Deutsche Forschungsgemeinschaft (DFG, German Research Foundation) under the German Excellence Initiative -- EXC 2121  Quantum Universe -- 390833306, by PUNCH4NFDI -- project number 460248186, as well as via the SciFM consortium (05D25GU4) funded by the German Federal Ministry of Research, Technology, and Space (BMFTR) in the ErUM-Data action plan. 
JB also acknowledges support via the Hamburg VISTA/VISOR --- Virtual Initiative for Science \& Technology in AI --- network.
This work has used the the Maxwell computational resources at Deutsches Elektronen-Synchrotron DESY, Hamburg, Germany, as well as the HPC-cluster Hummel-2 at University of Hamburg. Hummel-2 was funded by Deutsche Forschungsgemeinschaft (DFG, German Research Foundation) -- 498394658.}

\bibliographystyle{iopart-num}
\bibliography{refs}

\providecommand{\newblock}{}
\begin{thebibliography}{10}
\expandafter\ifx\csname url\endcsname\relax
  \def\url#1{{\tt #1}}\fi
\expandafter\ifx\csname urlprefix\endcsname\relax\def\urlprefix{URL }\fi
\providecommand{\eprint}[2][]{\url{#2}}

\bibitem{Agostinelli:2002hh}
Agostinelli S {\em et~al.\/} (GEANT4) 2003 {\em Nucl. Instrum. Meth.\/} {\bf
  A506} 250--303

\bibitem{2019}
Albrecht J {\em et~al.\/} 2019 {\em Computing and Software for Big Science\/}
  {\bf 3} ISSN 2510-2044
  \urlprefix\url{http://dx.doi.org/10.1007/s41781-018-0018-8}

\bibitem{Boehnlein:2803119}
Boehnlein A, Biscarat C, Bressan A, Britton D, Bolton R, Gaede F, Grandi C,
  Hernandez F, Kuhr T, Merino G, Simon F and Watts G 2022 {HL-LHC Software and
  Computing Review Panel, 2nd Report} Tech. rep. CERN Geneva
  \urlprefix\url{https://cds.cern.ch/record/2803119}

\bibitem{Paganini:2017hrr}
Paganini M, de~Oliveira L and Nachman B 2018 {\em Phys. Rev. Lett.\/} {\bf 120}
  042003 (\textit{Preprint} \eprint{https://arxiv.org/abs/1705.02355})

\bibitem{Paganini:2017dwg}
Paganini M, de~Oliveira L and Nachman B 2018 {\em Phys. Rev. D\/} {\bf 97}
  014021 (\textit{Preprint} \eprint{https://arxiv.org/abs/1712.10321})

\bibitem{deOliveira:2017rwa}
de~Oliveira L, Paganini M and Nachman B 2018 {\em J. Phys. Conf. Ser.\/} {\bf
  1085} 042017 (\textit{Preprint} \eprint{https://arxiv.org/abs/1711.08813})

\bibitem{Erdmann:2018kuh}
Erdmann M, Geiger L, Glombitza J and Schmidt D 2018 {\em Comput. Softw. Big
  Sci.\/} {\bf 2} 4 (\textit{Preprint}
  \eprint{https://arxiv.org/abs/1802.03325})

\bibitem{Erdmann:2018jxd}
Erdmann M, Glombitza J and Quast T 2019 {\em Comput. Softw. Big Sci.\/} {\bf 3}
  4 (\textit{Preprint} \eprint{https://arxiv.org/abs/1807.01954})

\bibitem{Musella:2018rdi}
Musella P and Pandolfi F 2018 {\em Comput. Softw. Big Sci.\/} {\bf 2} 8
  (\textit{Preprint} \eprint{https://arxiv.org/abs/1805.00850})

\bibitem{Belayneh:2019vyx}
Belayneh D {\em et~al.\/} 2020 {\em Eur. Phys. J. C\/} {\bf 80} 688
  (\textit{Preprint} \eprint{https://arxiv.org/abs/1912.06794})

\bibitem{Butter:2020qhk}
Butter A, Diefenbacher S, Kasieczka G, Nachman B and Plehn T 2021 {\em SciPost
  Phys.\/} {\bf 10} 139 (\textit{Preprint}
  \eprint{https://arxiv.org/abs/2008.06545})

\bibitem{ATLAS:2020}
{ATLAS collaboration} 2020-11 {Fast simulation of the ATLAS calorimeter system
  with Generative Adversarial Networks} techreport ATL-SOFT-PUB-2020-006 CERN
  (\textit{Preprint} \eprint{https://cds.cern.ch/record/2746032})

\bibitem{Ghosh:2020kkt}
{ATLAS Collaboration} 2020 {\em J. Phys. Conf. Ser.\/} {\bf 1525} 012077

\bibitem{ATLAS:2021pzo}
{ATLAS Collaboration} 2022 {\em Comput. Softw. Big Sci.\/} {\bf 6} 7
  (\textit{Preprint} \eprint{https://arxiv.org/abs/2109.02551})

\bibitem{ATLAS:2022jhk}
{ATLAS Collaboration} 2024 {\em Comput. Softw. Big Sci.\/} {\bf 8} 7
  (\textit{Preprint} \eprint{https://arxiv.org/abs/2210.06204})

\bibitem{FaucciGiannelli:2023fow}
Faucci~Giannelli M and Zhang R 2024 {\em Eur. Phys. J. Plus\/} {\bf 139} 597
  (\textit{Preprint} \eprint{https://arxiv.org/abs/2309.06515})

\bibitem{Dogru:2024gpk}
Simsek E, Isildak B, Dogru A, Aydogan R, Bayrak A~B and Ertekin S 2024 {\em
  PTEP\/} {\bf 2024} 083C01 (\textit{Preprint}
  \eprint{https://arxiv.org/abs/2401.02248})

\bibitem{Cresswell:2022tof}
Cresswell J~C, Ross B~L, Loaiza-Ganem G, Reyes-Gonzalez H, Letizia M and
  Caterini A~L 2022-11 {CaloMan: Fast generation of calorimeter showers with
  density estimation on learned manifolds} {\em {36th Conference on Neural
  Information Processing Systems}: {Workshop on Machine Learning and the
  Physical Sciences}\/} (\textit{Preprint}
  \eprint{https://arxiv.org/abs/2211.15380})

\bibitem{Hoque:2023zjt}
Hoque S, Jia H, Abhishek A, Fadaie M, Toledo-Marín J~Q, Vale T, Melko R~G,
  Swiatlowski M and Fedorko W~T 2024 {\em Eur. Phys. J. C\/} {\bf 84} 1244
  (\textit{Preprint} \eprint{https://arxiv.org/abs/2312.03179})

\bibitem{Liu:2024kvv}
Liu Q, Shimmin C, Liu X, Shlizerman E, Li S and Hsu S~C 2024-05 {Calo-VQ:
  Vector-Quantized Two-Stage Generative Model in Calorimeter Simulation}
  (\textit{Preprint} \eprint{https://arxiv.org/abs/2405.06605})

\bibitem{Krause:2021ilc}
Krause C and Shih D 2023 {\em Phys. Rev. D\/} {\bf 107} 113003
  (\textit{Preprint} \eprint{https://arxiv.org/abs/2106.05285})

\bibitem{Krause:2021wez}
Krause C and Shih D 2023 {\em Phys. Rev. D\/} {\bf 107} 113004
  (\textit{Preprint} \eprint{https://arxiv.org/abs/2110.11377})

\bibitem{Schnake:2022}
Schnake S, Krücker D and Borras K 2022 Generating calorimeter showers as point
  clouds {\em {36th Conference on Neural Information Processing Systems}:
  {Workshop on Machine Learning and the Physical Sciences}\/}
  (\textit{Preprint}
  \eprint{https://ml4physicalsciences.github.io/2022/files/NeurIPS\_ML4PS\_2022\_77.pdf})

\bibitem{Krause:2022jna}
Krause C, Pang I and Shih D 2024 {\em SciPost Phys.\/} {\bf 16} 126
  (\textit{Preprint} \eprint{https://arxiv.org/abs/2210.14245})

\bibitem{Xu:2023xdc}
Xu A, Han S, Ju X and Wang H 2024 {\em JINST\/} {\bf 19} P02003
  (\textit{Preprint} \eprint{https://arxiv.org/abs/2303.10148})

\bibitem{Buckley:2023daw}
Buckley M~R, Krause C, Pang I and Shih D 2024 {\em Phys. Rev. D\/} {\bf 109}
  033006 (\textit{Preprint} \eprint{https://arxiv.org/abs/2305.11934})

\bibitem{Pang:2023wfx}
Pang I, Shih D and Raine J~A 2024 {\em Phys. Rev. D\/} {\bf 109} 092009
  (\textit{Preprint} \eprint{https://arxiv.org/abs/2308.11700})

\bibitem{Ernst:2023qvn}
Ernst F, Favaro L, Krause C, Plehn T and Shih D 2025 {\em SciPost Phys.\/} {\bf
  18} 081 (\textit{Preprint} \eprint{https://arxiv.org/abs/2312.09290})

\bibitem{Schnake:2024mip}
Schnake S, Krücker D and Borras K 2024-03 {CaloPointFlow II Generating
  Calorimeter Showers as Point Clouds} (\textit{Preprint}
  \eprint{https://arxiv.org/abs/2403.15782})

\bibitem{Du:2024gbp}
Du H, Krause C, Mikuni V, Nachman B, Pang I and Shih D 2025 {\em Phys. Rev.
  D\/} {\bf 111} 076004 (\textit{Preprint}
  \eprint{https://arxiv.org/abs/2404.18992})

\bibitem{Majerz:2025ykn}
Majerz E, Dzwinel W and Kitowski J 2025-12 {Inverse Autoregressive Flows for
  Zero Degree Calorimeter fast simulation} {\em {39th Annual Conference on
  Neural Information Processing Systems}: {Includes Machine Learning and the
  Physical Sciences (ML4PS)}\/} (\textit{Preprint}
  \eprint{https://arxiv.org/abs/2512.20346})

\bibitem{Mikuni:2022xry}
Mikuni V and Nachman B 2022 {\em Phys. Rev. D\/} {\bf 106} 092009
  (\textit{Preprint} \eprint{https://arxiv.org/abs/2206.11898})

\bibitem{Acosta:2023zik}
Acosta F~T, Mikuni V, Nachman B, Arratia M, Karki B, Milton R, Karande P and
  Angerami A 2024 {\em JINST\/} {\bf 19} P05003 (\textit{Preprint}
  \eprint{https://arxiv.org/abs/2307.04780})

\bibitem{Mikuni:2023tqg}
Mikuni V and Nachman B 2024 {\em JINST\/} {\bf 19} P02001 (\textit{Preprint}
  \eprint{https://arxiv.org/abs/2308.03847})

\bibitem{Amram:2023onf}
Amram O and Pedro K 2023 {\em Phys. Rev. D\/} {\bf 108} 072014
  (\textit{Preprint} \eprint{https://arxiv.org/abs/2308.03876})

\bibitem{Jiang:2024ohg}
Jiang C, Qian S and Qu H 2025 {\em SciPost Phys.\/} {\bf 18} 195
  (\textit{Preprint} \eprint{https://arxiv.org/abs/2401.13162})

\bibitem{Kobylianskii:2024ijw}
Kobylianskii D, Soybelman N, Dreyer E and Gross E 2024 {\em Phys. Rev. D\/}
  {\bf 110} 072003 (\textit{Preprint}
  \eprint{https://arxiv.org/abs/2402.11575})

\bibitem{Jiang:2024bwr}
Jiang C, Qian S and Qu H 2024-04 {BUFF: Boosted Decision Tree based Ultra-Fast
  Flow matching} (\textit{Preprint} \eprint{https://arxiv.org/abs/2404.18219})

\bibitem{Favaro:2024rle}
Favaro L, Ore A, Schweitzer S~P and Plehn T 2025 {\em SciPost Phys.\/} {\bf 18}
  088 (\textit{Preprint} \eprint{https://arxiv.org/abs/2405.09629})

\bibitem{hildebrandt2026brickscompositionalneuralmarkov}
Hildebrandt R, Kourlitis E, Hashemi B, Bünstorf M, Meyer T, Boskov N, Kagan M,
  Rosenbaum D, Ganguly S and Heinrich L 2026 Bricks: Compositional neural
  markov kernels for zero-shot radiation-matter simulation (\textit{Preprint}
  \eprint{2605.06591}) \urlprefix\url{https://arxiv.org/abs/2605.06591}

\bibitem{li2026nestedgptvariablemultiplicitypartonshowers}
Li W, Shao D~Y, Shi H~Z and Sun Y~X 2026 Nested-gpt for variable-multiplicity
  parton showers: A case study in the resummation of non-global logarithms
  (\textit{Preprint} \eprint{2605.18360})
  \urlprefix\url{https://arxiv.org/abs/2605.18360}

\bibitem{bommasani2022opportunitiesrisksfoundationmodels}
Bommasani R {\em et~al.\/} 2022 On the opportunities and risks of foundation
  models (\textit{Preprint} \eprint{2108.07258})
  \urlprefix\url{https://arxiv.org/abs/2108.07258}

\bibitem{Hallin:2025ywf}
Hallin A 2025 {Foundation models for high-energy physics} {\em {2nd European AI
  for Fundamental Physics Conference}\/} (\textit{Preprint}
  \eprint{2509.21434})

\bibitem{Birk:2024knn}
Birk J, Hallin A and Kasieczka G 2024 {\em Mach. Learn. Sci. Tech.\/} {\bf 5}
  035031 (\textit{Preprint} \eprint{2403.05618})

\bibitem{Radford2018ImprovingLU}
Radford A, Narasimhan K, Salimans T and Sutskever I 2018 Improving language
  understanding by generative pre-training
  \urlprefix\url{https://cdn.openai.com/research-covers/language-unsupervised/language_understanding_paper.pdf}

\bibitem{oord2018neural}
van~den Oord A, Vinyals O and Kavukcuoglu K 2018 Neural discrete representation
  learning (\textit{Preprint} \eprint{1711.00937})

\bibitem{bao2022beit}
Bao H, Dong L, Piao S and Wei F 2022 {BEiT: BERT Pre-Training of Image
  Transformers} (\textit{Preprint} \eprint{2106.08254})

\bibitem{huh2023straightening}
Huh M, Cheung B, Agrawal P and Isola P 2023 Straightening out the
  straight-through estimator: Overcoming optimization challenges in vector
  quantized networks (\textit{Preprint} \eprint{2305.08842})

\bibitem{Birk:2025wai}
Birk J, Gaede F, Hallin A, Kasieczka G, Mozzanica M and Rose H 2025 {\em
  JINST\/} {\bf 20} P07007 (\textit{Preprint} \eprint{2501.05534})

\bibitem{cardonagiraldo2026generalizablefoundationmodelscalorimetry}
Cardona-Giraldo C, Fanelli C, Giroux J, Granger C, Nachman B and Sabin G 2026
  Generalizable foundation models for calorimetry via mixtures-of-experts and
  parameter efficient fine tuning (\textit{Preprint} \eprint{2603.28804})
  \urlprefix\url{https://arxiv.org/abs/2603.28804}

\bibitem{copet2024simplecontrollablemusicgeneration}
Copet J, Kreuk F, Gat I, Remez T, Kant D, Synnaeve G, Adi Y and Défossez A
  2024 Simple and controllable music generation (\textit{Preprint}
  \eprint{2306.05284}) \urlprefix\url{https://arxiv.org/abs/2306.05284}

\bibitem{Buss:2026yrf}
Buss T, Day-Hall H, Gaede F, Kasieczka G and Kr{\"u}ger K 2026 {AllShowers: One
  model for all calorimeter showers} (\textit{Preprint} \eprint{2601.11716})

\bibitem{Buhmann:2020pmy}
Buhmann E, Diefenbacher S, Eren E, Gaede F, Kasieczka G, Korol A and Kr{\"u}ger
  K 2021 {\em Comput. Softw. Big Sci.\/} {\bf 5} 13 (\textit{Preprint}
  \eprint{2005.05334})

\bibitem{Frank_2014}
Frank M, Gaede F, Grefe C and Mato P 2014 {\em Journal of Physics: Conference
  Series\/} {\bf 513} 022010
  \urlprefix\url{https://doi.org/10.1088/1742-6596/513/2/022010}

\bibitem{ILDConceptGroup:2020sfq}
Abramowicz H {\em et~al.\/} (ILD Concept Group) 2020 {International Large
  Detector: Interim Design Report} (\textit{Preprint} \eprint{2003.01116})

\bibitem{Suehara:2018mqk}
Suehara T {\em et~al.\/} 2018 {\em JINST\/} {\bf 13} C03015 (\textit{Preprint}
  \eprint{1801.02024})

\bibitem{ba2016layernormalization}
Ba J~L, Kiros J~R and Hinton G~E 2016 Layer normalization (\textit{Preprint}
  \eprint{1607.06450}) \urlprefix\url{https://arxiv.org/abs/1607.06450}

\bibitem{Bishop:1994}
Bishop C 1994 Mixture density networks WorkingPaper 4288 Aston University
  copyright {\textcopyright} 1994, Christopher M. Bishop. This work is licensed
  under a Creative Commons Attribution-NonCommercial-NoDerivatives 4.0
  International License (https://creativecommons.org/licenses/by-nc-nd/4.0/).

\bibitem{SU2024127063}
Su J, Ahmed M, Lu Y, Pan S, Bo W and Liu Y 2024 {\em Neurocomputing\/} {\bf
  568} 127063 ISSN 0925-2312
  \urlprefix\url{https://www.sciencedirect.com/science/article/pii/S0925231223011864}

\bibitem{shazeer2019fast}
Shazeer N 2019 Fast transformer decoding: One writer, en heads
  (\textit{Preprint} \eprint{1911.02150})

\bibitem{dao2022flashattention}
Dao T, Fu D~Y, Ermon S, Rudra A and R{\'e} C 2022 Flashattention: Fast and
  memory-efficient exact attention with io-awareness {\em Advances in Neural
  Information Processing Systems\/} vol~35 pp 10488--10500

\bibitem{dao2023flashattention2}
Dao T 2024 Flashattention-2: Faster attention with better parallelism and work
  partitioning {\em The Twelfth International Conference on Learning
  Representations\/}

\bibitem{NEURIPS2019_90fd4f88}
Zhang M, Lucas J, Ba J and Hinton G~E 2019 Lookahead optimizer: k steps
  forward, 1 step back {\em Advances in Neural Information Processing
  Systems\/} vol~32 ed Wallach H, Larochelle H, Beygelzimer A,
  d\textquotesingle Alch\'{e}-Buc F, Fox E and Garnett R (Curran Associates,
  Inc.)
  \urlprefix\url{https://proceedings.neurips.cc/paper_files/paper/2019/file/90fd4f88f588ae64038134f1eeaa023f-Paper.pdf}

\bibitem{Liu2020On}
Liu L, Jiang H, He P, Chen W, Liu X, Gao J and Han J 2020 On the variance of
  the adaptive learning rate and beyond {\em International Conference on
  Learning Representations\/}
  \urlprefix\url{https://openreview.net/forum?id=rkgz2aEKDr}

\bibitem{tarvainen2018meanteachersbetterrole}
Tarvainen A and Valpola H 2018 Mean teachers are better role models:
  Weight-averaged consistency targets improve semi-supervised deep learning
  results (\textit{Preprint} \eprint{1703.01780})
  \urlprefix\url{https://arxiv.org/abs/1703.01780}

\bibitem{Heneka:2025fpe}
Heneka C, Nieser F, Ore A, Plehn T and Schiller D 2026 {\em SciPost Phys.\/}
  {\bf 20} 070 (\textit{Preprint} \eprint{2506.14757})

\bibitem{Buhmann_2023}
Buhmann E, Diefenbacher S, Eren E, Gaede F, Kasicezka G, Korol A, Korcari W,
  Krüger K and McKeown P 2023 {\em Journal of Instrumentation\/} {\bf 18}
  P11025 ISSN 1748-0221
  \urlprefix\url{http://dx.doi.org/10.1088/1748-0221/18/11/P11025}

\end{thebibliography}

\end{document}